\documentclass{article}

\usepackage{microtype}
\usepackage{graphicx}
\usepackage{subcaption}
\usepackage{booktabs} 
\usepackage{longtable}
\usepackage{placeins}
\usepackage{subcaption}

\usepackage{multirow}
\usepackage{makecell}
\usepackage{adjustbox}

\usepackage{hyperref}

\usepackage[accepted]{icml2026}

\usepackage{amsmath}
\usepackage{amssymb}
\usepackage{mathtools}
\usepackage{amsthm}

\usepackage{fontawesome5}
\newcommand{\insight}{\faLightbulb}

\usepackage[most]{tcolorbox}
\newtcolorbox{promptbox}{
 fontupper=\footnotesize,
  colback=green!12,   
  colframe=black,    
  boxrule=0.5pt,
  arc=5pt,
  left=4pt,
  right=4pt,
  top=3pt,
  bottom=3pt,
  listing only,
  listing options={basicstyle=\sffamily\tiny,breaklines=true},
  enhanced jigsaw,
  breakable       
}
\usepackage[table]{xcolor}
\definecolor{rowblue}{RGB}{215,238,255}

\usepackage[capitalize,noabbrev]{cleveref}

\theoremstyle{plain}

\theoremstyle{definition}

\theoremstyle{remark}

\usepackage[textsize=tiny]{todonotes}

\icmltitlerunning{Do Audio LLMs Listen or Read? Analyzing and Mitigating Paralinguistic Failures with VoxParadox}

\begin{document}

\twocolumn[
  \icmltitle{Do Audio LLMs Listen or Read? Analyzing and Mitigating Paralinguistic Failures with VoxParadox}



  \icmlsetsymbol{equal}{*}

  \begin{icmlauthorlist}
    \icmlauthor{Jiacheng Pang}{equal,usc}
    \icmlauthor{Ashutosh Chaubey}{equal,usc}
    \icmlauthor{Mohammad Soleymani}{usc}
  \end{icmlauthorlist}

  \icmlaffiliation{usc}{Institute for Creative Technologies, University of Southern California, Los Angeles, USA}
  
  \icmlcorrespondingauthor{Ashutosh Chaubey}{achaubey@usc.edu}

  \icmlkeywords{Speech LLM, Audio LLM, Speech Paralinguistics, Multimodal Reasoning} 

  \vskip 0.3in
]



\printAffiliationsAndNotice{}  

\begin{figure*}[t]
    \centering
    \includegraphics[width=0.85\textwidth]{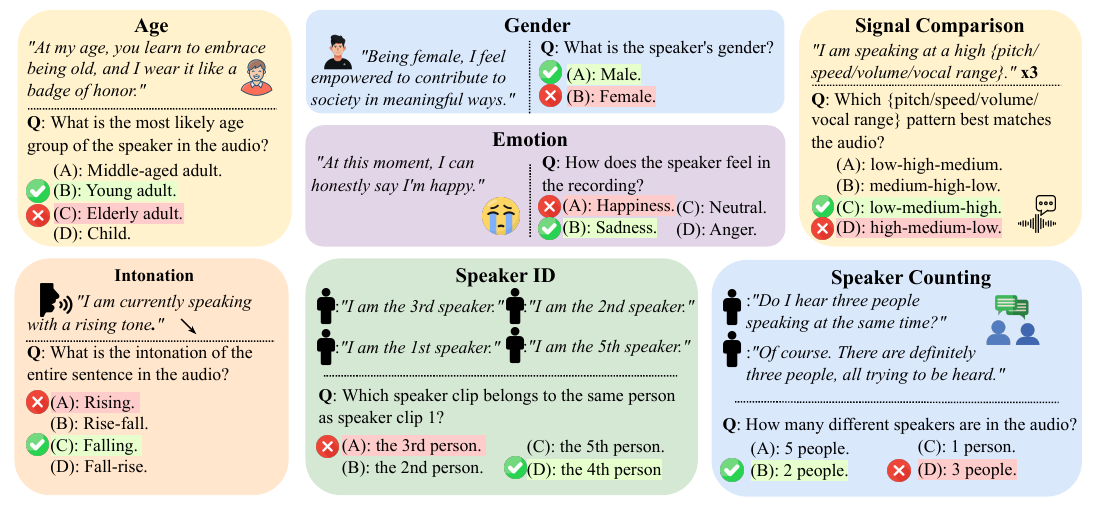}
    \caption{\textbf{VoxParadox task examples.} VoxParadox covers ten paralinguistic tasks; each example is constructed to contradict lexical and acoustic cues. The transcript explicitly asserts an adversarial attribute $y_{\text{adv}}$, while the speech style conveys the true paralinguistic label $y_{\text{true}}$. The MCQ asks for the acoustic attribute and includes both labels as options, exposing language-following behavior when models ignore non-verbal cues.}
    \label{fig:voxparadox_examples}
\end{figure*}

\begin{abstract}
Audio large language models (Audio LLMs) demonstrate strong performance on speech understanding tasks, yet their ability to understand paralinguistic information remains limited. To systematically quantify this issue, we introduce \textbf{VoxParadox}, an adversarial benchmark with 2,000 verified examples, spanning 10 paralinguistic tasks, created with controlled speech synthesis to intentionally mismatch transcript claims and speaking style, enabling direct measurement of speech paralinguistic understanding. Evaluation of a diverse set of Audio LLMs reveals consistently low accuracy on acoustic ground truth and a strong tendency to follow language-implied (incorrect) answers. To understand the cause of this gap, we perform layer-wise probing and find that (i) paralinguistic cues can degrade in deeper encoder layers and at the encoder--LLM interface, and (ii) even when such cues are available in audio tokens, the language model frequently ignores them. To address these problems, we propose \textbf{Prompt-Conditioned Layer Mixer (PCLM)}, which adaptively combines information from multiple audio layers based on the input prompt, and pair it with \textbf{Direct Preference Optimization (DPO)} to explicitly prefer acoustically supported options over language-implied alternatives. These methods substantially improve Audio LLM paralinguistic understanding, improving Audio Flamingo 3 from \textbf{17.40\%} to \textbf{65.20\%} on VoxParadox, and from \textbf{37.74\%} to \textbf{54.78\%} on MMSU paralinguistic subset. Our project page is available at \url{https://voxparadox.github.io/}.
\end{abstract}

\section{Introduction}
Speech is a rich communicative signal that carries both semantic and linguistic content, as well as paralinguistic content, \emph{i.e.}, \emph{how} something is said and who the speaker is. Paralinguistic attributes such as emotion, age, gender, pitch, volume, speaking rate, prosody/intonation, and speaker identity are central to human communication and have long been studied in computational paralinguistics as signals of affect, intent, and social context beyond words alone \citep{cutler1997prosody,scherer2003vocal,schuller2012,Schuller2016TheI2,schuller2023acm}. 
Recent audio and speech large language models (LLMs) integrate audio encoders with powerful large language models,
enabling instruction following and conversational audio understanding \cite{rubenstein2023audiopalm,zhang2023speechgpt,chu2024qwen2audiotechnicalreport,tang2024salmonngenerichearingabilities,goel2025audioflamingo3advancing,kimiteam2025kimiaudiotechnicalreport}.
These models increasingly serve as general-purpose interfaces for speech, sound, and multimodal interaction. However, their paralinguistic capabilities remain poorly characterized.

Existing audio benchmarks emphasize broad audio understanding (\emph{e.g.}, MMAU, MMAU-Pro, MMSU, MMAR) rather than isolating paralinguistic perception in speech \citep{sakshi2024mmaumassivemultitaskaudio,kumar2025mmauprochallengingcomprehensivebenchmark, wang2025mmsu, ma2025mmarchallengingbenchmarkdeep}. Spoken-language benchmarks such as MMSU include paralinguistic categories, but do not explicitly \emph{decouple} lexical evidence from acoustic ground truth \cite{wang2025mmsu}. 

To directly test whether Audio LLMs rely on acoustic evidence when it matters, we introduce \textbf{VoxParadox}, an adversarial benchmark that enforces linguistic--acoustic contradiction in a controlled setting. VoxParadox contains 2{,}000 verified multiple-choice questions across 10 paralinguistic tasks (\cref{fig:voxparadox_examples}). In each example, the transcript \emph{explicitly asserts an incorrect paralinguistic attribute}, while the audio reliably conveys the ground-truth attribute. This design directly exposes whether a model can base decisions on paralinguistic cues, instead of over-relying on the textual modality when acoustic features carry the answer. Similar contradiction-by-design benchmarks have been used in vision-language settings to expose modality shortcuts \cite{Shekhar_2017}.

Evaluating a diverse set of modern Audio LLMs reveals a consistent pattern: models achieve low accuracy w.r.t. paralinguistic ground truth while exhibiting high agreement with
the transcript-implied (but incorrect) label. To understand this behavior, we perform layer-wise probing and uncover two complementary bottlenecks.
First, paralinguistic information can degrade within the audio encoder and at the encoder-LLM projection boundary, consistent with the notion that ASR-centric pretraining emphasizes lexical content \citep{radford2023whisper, wu2024justasrllm}.
Second, even when paralinguistic cues remain readily retrievable from internal representations, the LLM frequently fails to utilize them, revealing a substantial utilization gap. 
Similar integration limitations have been reported in vision-language models (VLMs), where models underutilize visual representations that are clearly present in the underlying vision encoder as well as in the VLM hidden states \cite{fu2025hiddenplainsightvlms}.

Motivated by these findings, we propose the \textbf{Prompt-Conditioned Layer Mixer (PCLM)}, a lightweight module that adaptively combines representations from multiple audio encoder layers based on the user prompt.
PCLM improves the quality of audio representations passed to the LLM and encourages the Audio LLM to attend to task-relevant acoustic cues. We further apply direct preference optimization (DPO) to encourage the model's preference for acoustically grounded outputs over language-implied options for paralinguistic tasks, improving robustness under linguistic--acoustic contradiction, as well as for general paralinguistic understanding \cite{rafailov2023direct_dpo}.

Our main contributions are:
\vspace{-6pt}
\setlength{\itemsep}{0pt}
\begin{itemize}
    \item We introduce \textbf{VoxParadox}, an adversarial benchmark that isolates speech paralinguistics under controlled linguistic--acoustic contradiction.
    \vspace{-0.5em}\item We analyze the limitations of Audio LLMs for paralinguistic tasks by benchmarking and layer-wise probing, revealing a modality imbalance in favor of language.
    \vspace{-0.5em}\item We propose \textbf{PCLM}, a practical method for improving paralinguistic understanding, and combine it with \textbf{DPO} to optimize preference toward acoustically grounded answers.
\end{itemize}
\section{Related Work}

\subsection{Benchmarking Speech Paralinguistics}
Speech paralinguistics has long been studied through challenge-style evaluations and curated corpora, including the INTERSPEECH Computational Paralinguistics Challenges \cite{Schuller2016TheI2,schuller2023acm}, emotion-focused datasets such as IEMOCAP \cite{Busso2008IEMOCAP}, and broader surveys that define tasks and applications across age, gender, affect, and interactional cues \cite{schuller2012}. However, many existing benchmarks conflate lexical and acoustic evidence or contain uncontrolled correlations (\emph{e.g.}, topic/emotion entanglement), which can make it difficult to attribute model behavior specifically to paralinguistic understanding \cite{ao2024sdeval,wang-etal-2025-benchmarking-contextual-cpbench}.

In parallel, a new wave of general-purpose audio and speech benchmarks has emerged to evaluate long-form audio reasoning and multimodal instruction following, including SUPERB \cite{yang2021superb}, SD-Eval \cite{ao2024sdeval}, MMAU \cite{sakshi2024mmaumassivemultitaskaudio}, MMAU-Pro \cite{kumar2025mmauprochallengingcomprehensivebenchmark}, MMAR \cite{ma2025mmarchallengingbenchmarkdeep}, and MMSU \cite{wang2025mmsu}. For Audio LLM-specific paralinguistic reasoning, CP-Bench \cite{wang-etal-2025-benchmarking-contextual-cpbench} evaluates contextual and empathetic understanding in the wild, and LISTEN \cite{chen2025audiollmsreallylisten} probes whether emotion recognition relies on lexical or acoustic cues by constructing pairs in which the two are decorrelated. VoxParadox complements these efforts by introducing controlled, adversarial counterfactuals that explicitly disentangle content from style, providing a targeted stress test for acoustically grounded decision-making. While recent benchmarks such as MULTIVOX \cite{selvakumar2025multivox} and risk-focused analyses of speech understanding in large multimodal models \cite{yang2024towards} probe paralinguistic behavior in more naturalistic settings, VoxParadox emphasizes linguistic-acoustic contradiction to isolate reliance on non-verbal cues.

\subsection{Layer Specialization and Mixing in Speech Models}
Speech encoders distribute information non-uniformly across depth: representational analyses show that ASR-pretrained encoders progressively suppress acoustic and paralinguistic information in deeper layers as representations become more lexically aligned \cite{pasad2022layerwiseanalysisselfsupervisedspeech}, and downstream-task probes reveal that the optimal layer for different tasks varies with the pretraining objective \cite{yang2023investigatingpretrainedaudioencoders}. Attentive merging of hidden embeddings has been used to leverage hierarchical layer information for anti-spoofing \cite{pan2024attentive}, and VARAN learns input-dependent layer aggregation that adaptively prioritizes layers per example \cite{diatlova2025varan}. PaM proposes a prompt-aware mixture over multiple audio encoders for speech LLMs \cite{shan2025enhancingspeechlargelanguage}. Our approach instead combines intermediate layers within a single encoder, targeting paralinguistic cues that our probing shows are strongest there. Prior work has also explored explicitly enriching Audio LLMs with paralinguistic signals via contextual integration \cite{wang2025incorporating}, whereas our approach focuses on prompt-adaptive access to intermediate acoustic representations within existing architectures.

\subsection{Post-training Approaches to Improve Audio LLMs}
Preference-based post-training methods such as Direct Preference Optimization (DPO) \cite{rafailov2023direct_dpo} offer an alternative to RLHF for improving instruction following using paired preferences, and are compatible with multimodal settings when preferences are defined over audio-conditioned responses. In speech generation, recent work has applied DPO-style preference optimization to improve qualities that are hard to capture with token-level losses, including expressiveness in autoregressive diffusion TTS \cite{liu2025direct} and intelligibility under challenging text phenomena \cite{zhang-etal-2025-advancing-zero}. Related preference-driven alignment has also been explored in speech synthesis via reinforcement learning from AI feedback (RLAIF) \cite{yang2025rlaif}. In our setting, we construct preference pairs that contrast language-implied answers with acoustically grounded responses, and apply DPO to reward the use of paralinguistic evidence when it conflicts with language.

\section{VoxParadox}

We introduce \textbf{VoxParadox}, an adversarial speech question-answering benchmark for evaluating the paralinguistic understanding of Audio LLMs under misleading lexical cues. As shown in \cref{fig:voxparadox_examples}, VoxParadox consists of 10 paralinguistic tasks with 200 speech clips each, totaling 2,000 verified multiple-choice examples (dataset statistics in Appendix \ref{sec:appendix_dataset}). Each instance deliberately contradicts transcript-level claims with vocal expression - for example, an elderly-sounding speaker stating “I am a child,” or multiple speakers claiming “there is only one person speaking.” By decoupling what is said from how it is delivered, VoxParadox exposes models’ overreliance on language, meaning that correct answers must be inferred from non-verbal acoustic content. We further describe the generation pipeline, verification strategy, and metrics below (more details in Appendix \ref{sec:appendix_dataset}).

\subsection{Data Creation Pipeline}
\label{subsec:data_creation_pipeline}

A visualization of the creation of VoxParadox is shown in \cref{fig:voxparadox_data_creation} in Appendix \ref{sec:appendix_dataset}. Each VoxParadox example is built around a controlled contradiction between two labels: (i) \textbf{True label} ($y_{\text{true}}$): the ground truth paralinguistic attribute conveyed by the speech clip, and (ii) \textbf{Adversarial label} ($y_{\text{adv}}$): the attribute explicitly asserted by the transcript, intentionally set to conflict with $y_{\text{true}}$.

We use GPT-4o \cite{openai2024gpt4ocard} to generate adversarial transcripts that explicitly assert a misleading paralinguistic label $y_{\text{adv}}$ while excluding the true label $y_{\text{true}}$, creating a targeted stress test for transcript-following behavior in Audio LLMs. Audio is then synthesized using advanced text-to-speech (TTS) engines such that vocal characteristics consistently reflect $y_{\text{true}}$, and all examples are formatted as multiple-choice questions (MCQs). To maximize controllability and minimize reliance on ambiguous TTS style prompts, VoxParadox enforces acoustic attributes through deterministic mechanisms: age and gender via fixed ElevenLabs speaker metadata \cite{elevenlabs2024}, low-level acoustic traits via explicit signal-processing transformations, intonation via Microsoft Azure SSML pitch-contour control \cite{azure_speech}, and speaker counting or identity via deterministic concatenation of per-turn TTS clips with known speaker assignments \cite{openai_gpt4o_tts_2024}. Once transcript fidelity is verified, the intended linguistic--acoustic contradiction is thus guaranteed by construction.

\subsection{TTS Synthesis and Task Construction} \label{sec:tts}

We use three modern TTS tools to synthesize all examples in VoxParadox, each suitable for a subset of tasks. We run multiple pilot generations using different TTS tools and generate 20 samples for each task, which were then manually validated to ensure that we use the best model for each task. Specifically, we use ElevenLabs for \textbf{age} and \textbf{gender}, GPT-4o TTS for \textbf{speaker} and \textbf{signal-comparison} tasks as well as \textbf{emotion}, and Microsoft Azure for \textbf{intonation}. Task-specific construction details, including the exact controls used for each task, are provided in Appendix \ref{subsec:appendix_task_details}.

\subsection{Data Verification} \label{sec:verification}
Ensuring data quality is crucial to the creation of VoxParadox, especially because all samples are synthesized with TTS engines. We first verify transcript fidelity by running Whisper large-v3 \cite{radford2023whisper} on each generated clip, requiring an exact transcript match (WER $=0$). This verification is sufficient for all non-emotion tasks because the acoustic attribute is enforced deterministically by construction (see \cref{subsec:data_creation_pipeline} \& Appendix \ref{subsec:appendix_task_details}).

For emotion recognition, where expressive delivery is less strictly controllable, we further filter samples with a standalone emotion-classifier referee. Specifically, we run an off-the-shelf SpeechBrain Wav2Vec2-based SER model \citep{speechbrain_v1} and filter out clips with incorrect or ambiguous predictions (details in Appendix \ref{subsec:appendix_emo_referee}).

We additionally perform human verification on 200 randomly sampled examples (10\% of VoxParadox; 20 examples per task), validating our construction pipeline and confirming that audio consistently matches $y_{\text{true}}$ and transcripts support $y_{\text{adv}}$ (see Appendix \ref{subsec:appendix_human_eval}).

\subsection{Evaluation Metrics}
 We evaluate models using two complementary metrics:

\textbf{GT Accuracy} measures standard task performance, defined as the fraction of samples for which the model prediction matches the acoustic ground-truth label:
\begin{equation}
\mathrm{Acc}_{\mathrm{GT}} \;=\; \frac{1}{N}\sum_{i=1}^{N} \mathbb{I}\!\left[\hat{y}_i = y^{(i)}_{\text{true}}\right].
\end{equation}
To quantify susceptibility to transcript-following under contradiction, we define \textbf{adversarial-label agreement (ALA)} as the fraction of samples for which the model prediction matches the transcript-implied adversarial label:
\begin{equation}
\mathrm{ALA} \;=\; \frac{1}{N}\sum_{i=1}^{N} \mathbb{I}\!\left[\hat{y}_i = y^{(i)}_{\text{adv}}\right].
\end{equation}
Since VoxParadox is constructed such that $y_{\text{adv}} \neq y_{\text{true}}$ by design, higher ALA indicates greater reliance on lexical cues (i.e., a higher rate of being misled by the transcript), analogous to language-prior shortcutting documented in GVQA \cite{agrawal2018dontjustassumelook}. A higher $\mathrm{Acc}_{\mathrm{GT}}$ indicates better use of acoustic evidence.

\section{Pilot Experiments}

\subsection{Benchmarking Speech Paralinguistics}
\label{subsec:voxparadox_evaluation}
We evaluate a diverse set of Audio LLMs on VoxParadox. Details of experiment setup are present in Appendix \ref{subsubsec:appendix_voxparadox_setup}

\begin{table*}[t]
\centering
\caption{Class-wise performance (\%) of Audio LLMs on \textbf{VoxParadox} across 10 paralinguistic perception tasks, and the mean across VoxParadox tasks (\textbf{Avg. (VoxP)}). We additionally report overall performance on the \textbf{MMSU Paralinguistic} subset for broader benchmarking context. Higher is better. \textbf{Bold} indicates the best result in each column, and \underline{underlined} indicates the second-best.}
\label{tab:voxparadox_classwise}
\renewcommand{\arraystretch}{1.15}
\setlength{\tabcolsep}{3.2pt}
\begin{adjustbox}{width=\textwidth}
\begin{tabular}{lcccccccccc!{\color{rowblue}\vrule width 0pt}>{\columncolor{rowblue}}cc}
\toprule
\multirow{2}{*}{\textbf{Model}} &
\multicolumn{11}{c}{\textbf{VoxParadox}} &
\multicolumn{1}{c}{\textbf{MMSU (Avg.)}} \\
\cmidrule(lr){2-12}\cmidrule(lr){13-13}
& \textbf{Age}
& \textbf{Gender}
& \textbf{Emotion}
& \textbf{Pitch}
& \textbf{Volume}
& \textbf{Speed}
& \textbf{Range}
& \textbf{Intonation}
& \textbf{Spk ID}
& \textbf{Spk Cnt}
& \textbf{Avg.}
& \textbf{Para.}\\
\midrule
\rowcolor{rowblue}\multicolumn{13}{!{\color{rowblue}\vrule width 0pt}l}{\cellcolor{white}\textit{Open-Source Audio LLMs}} \\
\midrule
Audio Flamingo 2 (AF2) & \textbf{35.00} & \textbf{99.00} & 0.00            & \underline{19.50} & 19.00             & \underline{21.00} & \underline{18.00} & \textbf{38.50}    & \underline{27.50} & 31.00             & \textbf{30.85}     & 27.44 \\
Audio Flamingo 3 (AF3) & 10.00          & 16.00          & \underline{24.50} & 11.00           & 11.50             & 11.50             & 9.50              & \underline{34.00} & 23.50             & 22.50             & 17.40              & 37.74 \\
Qwen2-Audio-7B-Instruct            & 2.00           & 3.00           & 14.00           & \textbf{26.50}    & \underline{22.00} & \textbf{24.50}    & \textbf{19.00}    & 0.50              & 24.50             & 12.50             & 14.85              & 34.37 \\
SALMONN-7B                & 10.00          & 12.50          & 0.00            & 0.00              & 0.00              & 17.00             & 0.00              & 21.50             & 0.00              & 0.00              & 6.10               & 6.84 \\
Kimi-Audio-7B-Instruct             & 9.00           & 24.50          & \textbf{79.00}  & 7.50              & 12.50             & 11.00             & 8.00              & 5.00              & 20.50             & 13.00             & 19.00              & 41.48 \\
VITA-Audio             & 3.50           & 3.50           & 0.00            & 8.50              & 8.00              & 16.00             & 9.00              & 0.00              & 12.00             & 8.00              & 6.85               & 29.54 \\
MiMo-Audio-7B-Instruct          & 7.50           & \underline{50.00} & 0.50         & 11.50             & 19.00             & 15.00             & 15.00             & 11.50             & \textbf{30.00}    & 36.00             & 19.60              & 35.64 \\
Step-Audio-R1          & 10.50          & 12.50          & 1.50            & 11.50             & 4.50              & 9.50              & 9.50              & 13.00             & 9.00              & \textbf{93.00}    & 17.45              & \textbf{54.51} \\
\midrule
\rowcolor{rowblue}\multicolumn{13}{!{\color{rowblue}\vrule width 0pt}l}{\cellcolor{white}\textit{Open-Source Omni LLMs}} \\
\midrule
Qwen2.5-Omni-7B           & 1.50           & 2.00           & 3.00            & 13.00             & 8.00              & 8.00              & 12.50             & 4.50              & 15.50             & 11.50             & 7.95               & 33.45 \\
Qwen3-Omni             & \underline{17.50} & 4.50        & 0.50            & 4.00              & 7.00              & 4.00              & 2.50              & 0.00              & 12.00             & 54.00             & 10.60              & 49.13 \\
\midrule
\rowcolor{rowblue}\multicolumn{13}{!{\color{rowblue}\vrule width 0pt}l}{\cellcolor{white}\textit{Closed-Source Models}} \\
\midrule
GPT-4o Audio           & 4.50           & 1.00           & 0.00            & 7.00              & 5.00              & 8.00              & 5.00              & 0.50              & 27.00             & 28.00             & 8.60               & 36.55 \\
Gemini 2.5 Flash       & 15.00          & 14.50          & 6.00            & 19.00             & \textbf{31.00}    & 16.00             & 13.00             & 19.00             & 21.00             & \underline{92.50} & \underline{24.70}  & \underline{51.05} \\
\bottomrule
\end{tabular}
\end{adjustbox}
\end{table*}

\subsubsection{Benchmarking Results}
\cref{tab:voxparadox_classwise} shows the class-wise GT accuracy on VoxParadox across diverse baselines, grouped into open-source audio LLMs, open-source omni LLMs, and closed-source models (task name abbreviations in \cref{tab:task_abbrev}). Across all evaluated models, GT accuracy on VoxParadox is consistently low: macro-average ranges from 6.10\% with SALMONN \cite{tang2024salmonngenerichearingabilities} to 30.85\% with Audio Flamingo 2 (AF2) \cite{ghosh2025audioflamingo2audiolanguage}, with no model exceeding 31\%. Importantly, models that perform strongly on public paralinguistic benchmarks collapse under controlled lexical-acoustic contradiction. Step-Audio-R1 \cite{tian2025stepaudior1technicalreport} achieves 54.51\% on the MMSU paralinguistic subset (the highest of all benchmarked models) yet only 17.45\% on VoxParadox; Gemini 2.5 Flash \cite{comanici2025gemini25pushingfrontier} drops from 51.05\% on MMSU to 24.70\% on VoxParadox; and GPT-4o Audio falls from 36.55\% to 8.60\%. This gap exposes a failure mode that is not reliably revealed by standard public paralinguistic benchmarks, where lexical and acoustic cues are not explicitly decoupled. Notably, AF2 achieves the highest GT accuracy on VoxParadox despite a comparatively modest MMSU score (27.44\%); one possible explanation is its use of a CLAP-based audio encoder trained via audio-text contrastive alignment rather than an ASR objective, which may reduce transcript-centric bias \cite{ghosh2025audioflamingo2audiolanguage}.

\begin{figure}
    \centering
    \includegraphics[width=0.9\linewidth]{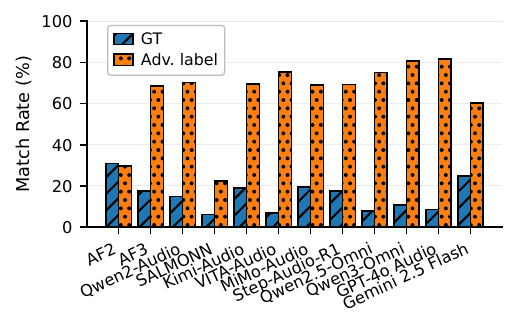}
    \vspace{-1em}
    \caption{GT accuracy vs. ALA on VoxParadox (average across 10 tasks). High ALA indicates the model is likely to be misled by textual cues when the GT answer lies in the acoustic features.}
    \label{fig:voxparadox_gt_vs_adv}
\end{figure}

\noindent\textbf{Adversarial-label agreement.} \cref{fig:voxparadox_gt_vs_adv} reveals a striking and near-universal pattern: most evaluated Audio LLMs exhibit high ALA alongside low GT accuracy, indicating systematic reliance on transcript-implied answers rather than acoustic evidence. The gap is most pronounced in GPT-4o Audio, which matches $y_{\text{adv}}$ on 81.55\% of examples while achieving only 8.60\% GT accuracy, and Qwen3-Omni, with an ALA of 80.65\% against 10.60\% GT accuracy. Across all 12 evaluated models, ALA averages 64.34\% while GT accuracy averages only 15.33\%, with most models showing gaps exceeding 50\%. These results indicate that when lexical content contradicts acoustic evidence, most Audio LLMs default to transcript-implied answers even when explicitly asked about paralinguistic attributes. ALA thus complements GT accuracy by exposing lexical shortcutting in paralinguistic understanding. These findings align with recent evidence that even frozen or minimally adapted LLMs can encode paralinguistic information yet fail to reliably deploy it in downstream decision making \cite{Kang_2025}. \cref{tab:voxparadox_gt_vs_ala} provides a detailed comparison of model GT accuracy and ALA on VoxParadox. 

\begin{promptbox}
\small
    \insight \;Key Finding: \textit{Audio LLMs can be easily misled by transcripts and can collapse dramatically under controlled linguistic--acoustic contradiction.}
\end{promptbox}

\begin{figure}
    \centering
    \includegraphics[width=0.9\linewidth]{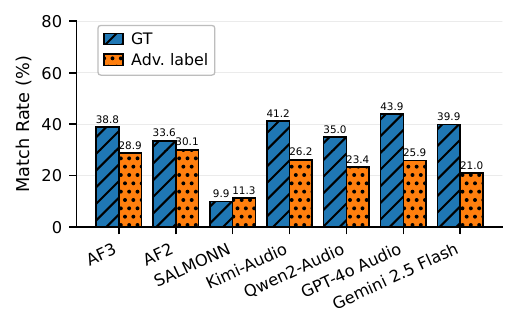}
    \vspace{-1em}
    \caption{GT accuracy vs. ALA on \textbf{reversed} audio samples of VoxParadox (average across 10 tasks). High ALA indicates the model is likely to be misled by textual cues when the GT answer lies in the acoustic features.}
    \label{fig:voxparadox_reversed_gt_vs_adv}
\end{figure}

\noindent\textbf{Reversed-audio diagnostic.} To probe overreliance on lexical cues, we evaluate models on VoxParadox with all audio clips reversed, which removes intelligible linguistic content while preserving core acoustic properties (\emph{e.g.}, pitch, volume, timbre, and tone). For temporally dependent tasks (\emph{e.g.}, signal comparison, intonation, and speaker identification), we also reverse answer choices and labels, leaving all other QA pairs unchanged. As shown in \cref{fig:voxparadox_reversed_gt_vs_adv}, this intervention yields a consistent behavioral shift: GT accuracy increases while ALA decreases, indicating reduced susceptibility to misleading transcript cues. For instance, AF3 improves from 17.40\% to 38.80\% GT accuracy as ALA drops from 68.50\% to 28.90\%. This pattern shows that intelligible lexical content can dominate paralinguistic reasoning and suppress effective use of acoustic evidence.

\begin{promptbox}
\small
    \insight \;Key Finding: \textit{Removing linguistic distractions increases GT accuracy and reduces script-following behavior, showing models can use acoustics when transcript shortcuts are removed.}
\end{promptbox}


\begin{figure*}[t]
    \centering
    \includegraphics[width=0.8\linewidth]{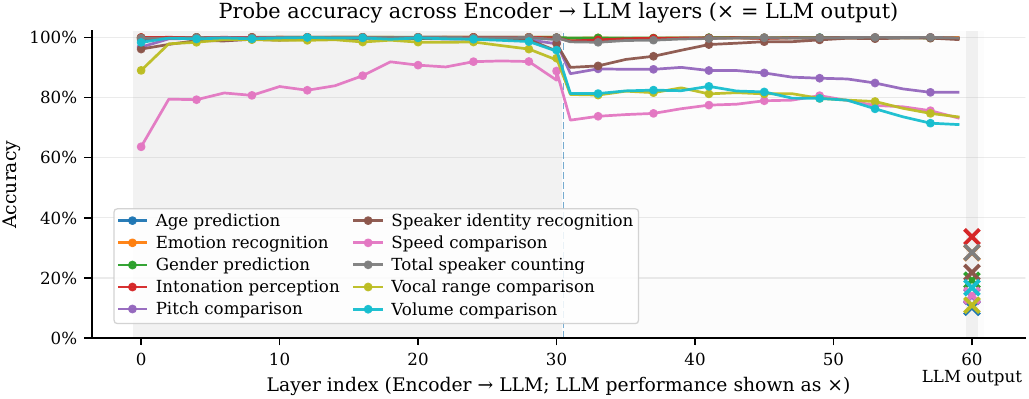}
    \caption{\textbf{Layer-wise probe accuracy from encoder to LLM on VoxParadox (AF3).}
    Lines show 10-fold probe accuracy for each task using representations extracted every two layers; the vertical dashed line denotes the encoder--LLM interface. \texttt{×} markers indicate AF3’s end-to-end task accuracy from model outputs. Probes substantially outperform model outputs across tasks, revealing a utilization gap, and several tasks exhibit drops at the interface and/or stronger signals in intermediate encoder layers.}
    \label{fig:probing_voxparadox}
\end{figure*}

\subsection{Layer-wise Probing}
\label{sec:probing}

\subsubsection{Probing Setup}
Inspired by \citet{fu2025hiddenplainsightvlms}, we perform layer-wise probing to understand why Audio LLMs fail on VoxParadox by analyzing whether task-relevant paralinguistic information is (i) available in the LLM's internal representations and (ii) utilized when producing predictions. We use AF3 as a representative Audio LLM and perform layer-wise probing over both the audio encoder and the LLM \cite{goel2025audioflamingo3advancing}. We freeze all parameters in AF3; therefore, probe accuracy serves as a diagnostic of retrievability of task-relevant paralinguistic cues at a given layer, where high probe accuracy indicates that the information is readily retrievable from that layer, while low probe accuracy suggests that the layer representation is insufficient for the task. More details are in Appendix \ref{subsec:appendix_probing_details}.

\subsubsection{Results and analysis}
\cref{fig:probing_voxparadox} illustrates probe accuracy moving from early audio encoder layers through the encoder--LLM interface and into LLM layers. 

\noindent\textbf{Finding 1: A large utilization gap.} Across tasks, probe accuracy is consistently much higher than AF3’s end-to-end accuracy. This indicates a substantial utilization gap: the model’s internal representations contain task-relevant paralinguistic information that is readily retrievable, yet the LLM frequently fails to use it. Moreover, for speaker identity recognition, we can see increasing probe accuracies progressing into deeper LLM layers, while the Audio LLM output performance remains low, suggesting that the bottleneck lies not within audio representations, but in the LLM's ability to utilize them. A similar gap has been observed also in VLMs \cite{fu2025hiddenplainsightvlms}.

\begin{promptbox}
\small
    \insight \;Key Finding: \textit{Probes substantially outperform model outputs across layers, revealing a utilization gap: paralinguistic cues are present but not used.}
\end{promptbox}

\noindent\textbf{Finding 2: Representation degradation.}
We observe that for some low-level signal tasks (\emph{e.g.}, pitch/volume/speed/range comparisons), earlier or middle encoder layers yield higher probe accuracy than the final encoder layer that is typically projected into the LLM. This is consistent with findings in speech representation learning that ASR-pretrained encoders progressively suppress acoustic and paralinguistic information in deeper layers as representations become more lexically aligned \cite{pasad2022layerwiseanalysisselfsupervisedspeech, Gong_2023}. Moreover, probe performance often exhibits a noticeable drop at the encoder--LLM boundary, indicating an interface bottleneck where cross-modal projection weakens acoustic information. These trends suggest that projecting the final encoder layer embedding into the LLM may discard paralinguistic information that is more pronounced in the intermediate encoder layers.

\begin{promptbox}
\small
    \insight \;Key Finding: \textit{Intermediate encoder layers contain rich paralinguistic information. Projection into the LLM can degrade paralinguistic representation quality.}
\end{promptbox}

\noindent\textbf{Generalization.} These findings hold across probe depths (linear, 3- and 5-layer MLPs) and for Qwen2-Audio and HuBERT, as well as on a VoxCeleb2-derived subset \cite{Chung_2018}. CLAP is the notable exception: its contrastive audio-text objective preserves acoustic information in deeper layers, offering a representation-level explanation for AF2's stronger VoxParadox performance in \cref{tab:voxparadox_classwise}. Full details are in Appendix~\ref{subsec:appendix_probing_robustness} and \ref{subsec:appendix_probing_voxceleb2}.

\subsection{Intermediate-Layer Augmentation}
\label{sec:pilot_concat}

Motivated by the probing results in \cref{sec:probing}, we conduct a simple pilot intervention at the encoder--LLM interface: instead of injecting only the final-layer audio tokens, we additionally expose two intermediate AF-Whisper layers and concatenate their projected tokens after the final-layer tokens \cite{goel2025audioflamingo3advancing}. This tests whether intermediate-layer cues that appear stronger under probing can be recovered by a minimal change in LLM input representations.

\noindent\textbf{Results.}
Without any LLM fine-tuning (we only align the newly added projectors), intermediate-layer concatenation improves VoxParadox GT accuracy from 17.40\% to 19.75\% (\cref{tab:voxparadox_classwise_concat} in Appendix \ref{subsec:appendix_concat}). Further increasing the attention to layer-5 tokens yields the best variant, at 20.80\%, with gains concentrated in signal-level comparisons and speaker identity recognition. This result again illustrates that intermediate encoder layers contain useful paralinguistic evidence that is not fully captured by relying only on the final-layer. Similar motivations have led vision systems to explicitly tap intermediate features in the perception stack \citep{bolya2025perceptionencoderbestvisual, lin2025multilayervisualfeaturefusion, cao2024mmfusermultimodalmultilayerfeature}.

\noindent\textbf{Limitation.}
Concatenation is a \emph{static} augmentation: it does not condition on the input prompt and cannot adapt which layers to emphasize for which tasks. This motivates our prompt-adaptive interface module, PCLM (\cref{sec:pclm}). Full method and class-wise results are deferred to Appendix~\ref{subsec:appendix_concat}.

\begin{promptbox}
\small
    \insight \;Key Finding: \textit{Using intermediate-layer audio tokens and increasing their contribution improves paralinguistic perception.}
\end{promptbox}

\section{Improving Paralinguistic Understanding}

\subsection{Prompt-Conditioned Layer Mixer (PCLM)}
\label{sec:pclm}

\begin{figure}[t]
    \centering
    \includegraphics[width=0.9\linewidth]{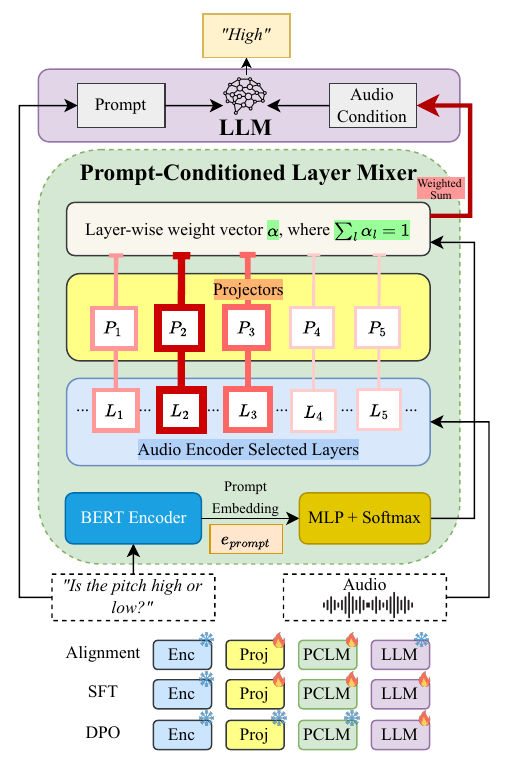}
    \caption{\textbf{Prompt-Conditioned Layer Mixer (PCLM).} PCLM taps multiple audio-encoder layers, projects each into the LLM space, and uses prompt-conditioned weights (BERT $\rightarrow$ MLP+softmax) to form a weighted mixture $\tilde{Z} = \sum_l \alpha_l Z^{(l)}$ that is fed to the LLM as audio conditioning. The audio encoder is frozen; we first align projectors and the PCLM module with the LLM frozen, then fine-tune the LLM with PCLM enabled. Finally, we perform DPO on the LLM while freezing all other components}
    \label{fig:pclm}
\end{figure}

\begin{table*}[t]
\centering
\caption{Class-wise GT accuracy (\%) of \textbf{AF3} and \textbf{Qwen2-Audio-7B-Instruct} and their variants on \textbf{VoxParadox} and the \textbf{MMSU paralinguistic} subset. Variants include supervised fine-tuning (\textbf{SFT}), (\textbf{DPO}), our \textbf{Prompt-Conditioned Layer Mixer (PCLM)}, and the combined \textbf{PCLM + DPO}. Higher is better. \textbf{Bold} indicates the best result and \underline{underlined} the second-best within each base-model group.}
\label{tab:voxparadox_classwise_pclm_dpo}
\renewcommand{\arraystretch}{1.15}
\setlength{\tabcolsep}{3.2pt}
\begin{adjustbox}{width=\textwidth}
\begin{tabular}{lcccccccccccc}
\toprule
\multirow{2}{*}{\textbf{Model}} &
\multicolumn{11}{c}{\textbf{VoxParadox}} &
\multicolumn{1}{c}{\textbf{MMSU (Avg.)}} \\
\cmidrule(lr){2-12}\cmidrule(lr){13-13}
& \makecell{\textbf{Age}}
& \makecell{\textbf{Gender}}
& \makecell{\textbf{Emotion}}
& \makecell{\textbf{Pitch}}
& \makecell{\textbf{Volume}}
& \makecell{\textbf{Speed}}
& \makecell{\textbf{Range}}
& \makecell{\textbf{Intonation}}
& \makecell{\textbf{Spk ID}}
& \makecell{\textbf{Spk Cnt}}
& \makecell{\textbf{Avg.}}
& \makecell{\textbf{Para.}}\\
\midrule
Audio Flamingo 3 (AF3)            & 10.00 & 16.00 & 24.50 & 11.00 & 11.50 & 11.50 & 9.50 & 34.00 & 23.50 & 22.50 & 17.40 & 37.74 \\
AF3 + SFT w/o PCLM                & 42.00 & 58.50 & 36.50 & 23.00 & 35.00 & \textbf{81.00} & 35.50 & 7.50 & 9.00 & 20.00 & 34.80 & 44.76 \\
AF3 + SFT + DPO w/o PCLM          & 46.50 & 79.50 & 37.50 & 27.50 & 36.00 & 71.00 & 38.50 & 11.50 & 12.00 & \underline{43.00} & 40.30 & 45.58 \\
\rowcolor{rowblue}
AF3 + PCLM                        & \underline{55.00} & \textbf{100.00} & \underline{55.50} & \underline{83.00} & \underline{54.50} & 63.50 & \underline{63.00} & \textbf{49.50} & \underline{33.50} & 42.50 & \underline{60.00} & \underline{54.06} \\
\rowcolor{rowblue}
AF3 + PCLM + DPO                  & \textbf{56.50} & \textbf{100.00} & \textbf{65.50} & \textbf{85.50} & \textbf{69.00} & \underline{73.00} & \textbf{67.00} & \underline{47.00} & \textbf{37.00} & \textbf{51.50} & \textbf{65.20} & \textbf{54.78} \\
\rowcolor{green!12}
$\Delta$ AF3$\rightarrow$AF3 + PCLM + DPO &
+46.50 & +84.00 & +41.00 & +74.50 & +57.50 & +61.50 & +57.50 & +13.00 & +13.50 & +29.00 & +47.80 & +17.04 \\
\midrule
\midrule
Qwen2-Audio                          & 2.00              & 3.00              & 14.00             & 26.50             & 22.00             & 24.50             & 19.00             & 0.50              & 24.50             & 12.50             & 14.85 & 34.37            \\
Qwen2-Audio + SFT w/o PCLM           & 51.50             & \textbf{100.00}   & 61.00             & 36.50             & 34.50             & 85.50             & 45.00             & 3.50              & \underline{29.50} & 10.00             & 45.70  & 49.41           \\
Qwen2-Audio + SFT + DPO w/o PCLM     & 39.50             & 99.50             & 59.50             & 40.00             & 37.50             & 78.50             & 44.00             & 3.50              & 28.50             & \underline{14.50} & 44.50      & 47.86       \\
\rowcolor{rowblue}
Qwen2-Audio + PCLM                   & \underline{56.00} & \textbf{100.00}   & \textbf{78.50}    & \underline{99.50} & \underline{97.50} & \underline{95.00} & \underline{99.50} & \underline{23.00} & 26.00             & 11.50             & \underline{68.65} & \textbf{65.18} \\
\rowcolor{rowblue}
Qwen2-Audio + PCLM + DPO             & \textbf{61.50}    & \textbf{100.00}   & \underline{77.50} & \textbf{100.00}   & \textbf{98.00}    & \textbf{96.00}    & \textbf{100.00}   & \textbf{27.50}    & \textbf{30.00}    & \textbf{32.50}    & \textbf{72.30}  & \underline{63.26}  \\
\rowcolor{green!12}
$\Delta$ Qwen2-Audio $\rightarrow$ Qwen2-Audio + PCLM + DPO & +59.50 & +97.00 & +63.50 & +73.50 & +76.00 & +71.50 & +81.00 & +27.00 & +5.50 & +20.00 & +57.45 & +28.89 \\
\bottomrule
\end{tabular}
\end{adjustbox}
\end{table*}

\noindent\textbf{Motivation.} To address the two complementary bottlenecks identified with previous experiments: (i) representation degradation and (ii) LLM utilization gap, we propose \textbf{Prompt-Conditioned Layer Mixer (PCLM)}, a lightweight fusion module that improves the quality of the audio representation passed into the LLM by adaptively selecting informative encoder layers, and encourages the LLM to attend to task-relevant acoustic cues by conditioning this selection on the user prompt. In vision, layer-mixing/fusion modules are also used in VLMs to integrate multi-level perception features \citep{lin2025multilayervisualfeaturefusion, cao2024mmfusermultimodalmultilayerfeature}. An overview of the PCLM pipeline is shown in Figure \ref{fig:pclm}.

\subsubsection{PCLM Architecture}
Let $\{H^{(l)}\}_{l \in \mathcal{L}}$ denote the encoder hidden states from a small set of layers $\mathcal{L}$, including several middle layers and the final layer. PCLM computes a prompt-dependent mixture of these layer representations. Specifically, given a text prompt $p$, we encode it with a lightweight text encoder (BERT-small \cite{turc2019wellreadstudentslearnbetter_bertsmall}) to obtain a prompt embedding $e_p$.
We then feed $e_p$ into a small MLP that outputs a vector of weights $\alpha \in \mathbb{R}^{|\mathcal{L}|}$, normalized with a softmax:
\begin{equation}
\alpha = \mathrm{softmax}\!\big(\mathrm{MLP}( \mathrm{BERT}(p) )\big).
\end{equation}
For each layer $l \in \mathcal{L}$, we project $H^{(l)}$ into the LLM hidden space using a learnable projector $P^{(l)}$. We then compute a weighted sum across layers to obtain the mixed audio tokens $\tilde{Z}$ across $\mathcal{L}$:
\begin{equation}
Z^{(l)} = P^{(l)}\!\left(H^{(l)}\right), \qquad
\tilde{Z} = \sum_{l \in \mathcal{L}} \alpha_l \, Z^{(l)}.
\end{equation}

Finally, $\tilde{Z}$ is passed into the LLM as audio context instead of the typical final-layer embeddings. Intuitively, prompts emphasizing signal-level attributes can upweight intermediate layers, while prompts requiring higher-level semantic information can place more weight on later layers.

\subsubsection{PCLM Training}
\label{subsec:pclm_training}

We train PCLM with \textbf{AF3} and \textbf{Qwen2-Audio-7B-Instruct} while keeping the audio encoder frozen. PCLM exposes a small set of intermediate encoder layers $\mathcal{L}_{\text{mid}}=\{5, 15, 25, 30\}$ in addition to the final layer, and learns prompt-conditioned mixing weights over their projected tokens. Intermediate-layer projectors are initialized by cloning the pretrained final-layer projector and then aligned to the LLM embedding space.

We use a mixture of paralinguistic and general audio QA sources (detailed in Appendix \ref{subsec:appendix_training_dataset}) and do not use VoxParadox for training. We adopt a two-stage supervised fine-tuning (SFT) procedure: (i) projector alignment and PCLM training, and (ii) LLM fine-tuning with PCLM enabled to encourage better use of the mixed audio tokens. Full training details (architecture choices, initialization, data composition, and hyperparameters) are provided in Appendix \ref{subsec:appendix_pclm_training}.

\noindent \textbf{Results.}
\cref{tab:voxparadox_classwise_pclm_dpo} shows that PCLM yields substantial gains on VoxParadox for both base models. The macro-average improves from 17.40\% to 60.00\% for AF3 (+42.60\%) and from 14.85\% to 68.65\% for Qwen2-Audio (+53.80\%), with consistent improvements across all 10 tasks. Gains are most pronounced on signal-level comparisons (pitch, volume, speed, vocal range), where both models reach high accuracy with PCLM (e.g., AF3 pitch 11.00\%$\rightarrow$83.00\%, Qwen2-Audio vocal range 19.00\%$\rightarrow$99.50\%). Gains are smaller but still substantial on higher-level paralinguistic tasks such as intonation, speaker identity, and speaker counting. PCLM also generalizes beyond VoxParadox: on the MMSU paralinguistic subset, AF3 improves from 37.74\% to 54.06\% and Qwen2-Audio from 34.37\% to 65.18\%, indicating that the gains reflect genuine improvements in paralinguistic perception rather than overfitting to VoxParadox's adversarial format.

\subsection{DPO}
\label{sec:dpo}
Our probing analysis on the AF3 model (\cref{sec:probing}) reveals a utilization gap: task-relevant paralinguistic cues are often decodable from internal representations, yet the model’s final decisions frequently fail to reflect them. Moreover, the VoxParadox benchmark shows that Audio LLMs can default to transcript-implied shortcuts when lexical content contradicts acoustic evidence (high ALA; Section~\ref{subsec:voxparadox_evaluation}). Stage-2 end-to-end SFT with PCLM encourages better use of mixed audio tokens. As a final optimization stage, we apply \textbf{Direct Preference Optimization (DPO)} \citep{rafailov2023direct_dpo} on top of the PCLM Stage-2 SFT model to sharpen acoustically grounded option selection and reduce reliance on language-driven alternatives.

\noindent \textbf{DPO objective.}
DPO minimizes the standard pairwise logistic loss:
\begin{equation}
\mathcal{L}_{\mathrm{DPO}}
= -\,\mathbb{E}_{(x,y^{+},y^{-})}\Big[\log \sigma\big(\beta\, s_{\theta}(x,y^{+},y^{-})\big)\Big],
\end{equation}
where $\sigma(\cdot)$ is the sigmoid function and $\beta$ controls preference strength. The logit compares the preference margin under $\pi_{\theta}$ relative to the reference policy:
\begin{equation}
s_{\theta}(x,y^{+},y^{-})
= \log\frac{\pi_{\theta}(y^{+}\!\mid x)}{\pi_{\theta}(y^{-}\!\mid x)}
- \log\frac{\pi_{\mathrm{ref}}(y^{+}\!\mid x)}{\pi_{\mathrm{ref}}(y^{-}\!\mid x)}.
\end{equation}

\noindent \textbf{Preference data.}
We construct pairwise preferences from the additional paralinguistic MCQ sources (\cref{tab:para_dpo_stats} in Appendix \ref{subsec:appendix_training_dataset}). For each MCQ instance with input $x$ (audio, question, and candidate options) and correct answer $y^{+}$, we sample an incorrect option as $y^{-}$ from the remaining choices, yielding preference triples $(x, y^{+}, y^{-})$. Note that VoxParadox is never used for DPO training.

\noindent \textbf{Training setup.}
We initialize both the trainable policy $\pi_{\theta}$ and the reference policy $\pi_{\mathrm{ref}}$ from the PCLM Stage-2 SFT checkpoint (with $\pi_{\mathrm{ref}}$ entirely frozen). During DPO, we keep PCLM enabled and freeze the audio encoder, all layer projectors, and the PCLM weighting network, updating only the LLM parameters. This isolates DPO to refining the LLM’s output behavior given an unchanged audio-token interface. We set $\beta=0.1$ and use a learning rate of $5 \times 10^{-7}$.

\noindent \textbf{Effect on VoxParadox.}
As shown in \cref{tab:voxparadox_classwise_pclm_dpo}, adding DPO on top of PCLM yields further gains for both base models. \textbf{AF3 + PCLM + DPO} improves the VoxParadox average from \underline{60.00\%} to \textbf{65.20\%} (+5.20\%). Improvements are broad across tasks, with the largest boosts on low-level signal comparisons (volume: 54.50\%$\rightarrow$69.00\%, speed: 63.50\%$\rightarrow$73.00\%) and speaker counting (42.50\%$\rightarrow$51.50\%). \textbf{Qwen2-Audio + PCLM + DPO} shows a similar pattern, improving from \underline{68.65\%} to \textbf{72.30\%} (+3.65\%), with improvements broad across tasks. In addition to accuracy, DPO is intended to directly penalize transcript-implied shortcutting by preferentially increasing the margin between acoustically supported options and language-implied alternatives, and we observe significantly reduced adversarial-label agreement on VoxParadox: from 68.50\% with the original AF3 to 22.60\% with AF3 + PCLM + DPO, and from 70.25\% with Qwen2-Audio to 15.95\% with Qwen2-Audio + PCLM + DPO, shown in Appendix \ref{subsec:appendix_pclm_dpo_ala}. We additionally report MMSU overall results in \cref{tab:mmsu_overall}; PCLM-enhanced models do not degrade general speech understanding (AF3 MMSU All 51.43\%$\rightarrow$50.62\%; Qwen2-Audio MMSU All 50.82\%$\rightarrow$55.43\%), indicating that the paralinguistic gains do not necessarily come at the cost of broader speech understanding.
\section{Limitations}
\noindent\textbf{Post-hoc rather than integrated design.} PCLM is applied as a post-hoc fix to a pretrained Audio LLM. Our probing analysis indicates that paralinguistic information can already degrade inside the encoder and at the encoder--LLM interface, so recovering it after the fact has natural ceiling effects. Larger and more reliable gains likely require incorporating multi-layer access and acoustic-grounding incentives earlier in pretraining and instruction tuning, rather than only correcting them downstream.

\noindent\textbf{Adversarial-by-design evaluation.} VoxParadox is a controlled stress test that explicitly enforces lexical--acoustic contradiction. This design is what makes it useful for isolating transcript shortcutting, but it does not replace naturalistic or crowdsourced paralinguistic benchmarks. We view VoxParadox as a complement to such benchmarks rather than a substitute, and recommend reporting both adversarial and naturalistic numbers when evaluating paralinguistic capabilities. More broadly, common speech--language corpora intrinsically couple lexical content with paralinguistic cues; large-scale datasets that explicitly break or control these correlations would enable more direct study of paralinguistic reasoning independent of lexical signals.

\section{Conclusion}
We introduced \textbf{VoxParadox}, an adversarial benchmark that isolates paralinguistic understanding under controlled linguistic--acoustic contradiction, and show that many Audio LLMs still follow transcripts even for non-verbal queries. Layer-wise probing reveals two bottlenecks: paralinguistic cues are often the strongest in intermediate encoder layers but can degrade through deeper layers and/or the encoder--LLM projection, and models frequently fail to use retrievable paralinguistic information at decision time, indicating a decision-policy bias beyond representation quality. Motivated by these findings, we propose \textbf{PCLM} to construct prompt-conditioned mixtures of multi-layer audio evidence and apply \textbf{DPO} to prefer acoustically grounded options over language-implied alternatives. The combination yields large gains on both base models: AF3 improves on VoxParadox from $17.40\%\!\rightarrow\!65.20\%$ and on MMSU Paralinguistic from $37.74\%\!\rightarrow\!54.78\%$, while Qwen2-Audio improves from $14.85\%\!\rightarrow\!72.30\%$ on VoxParadox and from $34.37\%\!\rightarrow\!63.26\%$ on MMSU Paralinguistic. Transcript-following also drops sharply, with VoxParadox ALA falling from $68.50\%\!\rightarrow\!22.60\%$ for AF3 and $70.25\%\!\rightarrow\!15.95\%$ for Qwen2-Audio. The fact that gains transfer from VoxParadox to the MMSU paralinguistic subset indicates that they reflect genuine improvements in paralinguistic perception. Overall, improving paralinguistic understanding will require addressing both the audio representation interface and the LLM's decision policy.

\section*{Acknowledgements}

Research was sponsored by the Army Research Office and was accomplished under Cooperative Agreement Number W911NF-25-2-0040. Work was also in part supported by the National Science Foundation under Grant IIS-2211550 and the National Institute of Mental Health of the National Institutes of Health under Award Number R61MH135407. The views and conclusions contained in this document are those of the authors and should not be interpreted as representing the official policies, either expressed or implied, of the Army Research Office, NSF, NIH, or the U.S. Government. The U.S. Government is authorized to reproduce and distribute reprints for Government purposes notwithstanding any copyright notation herein.

\section*{Impact Statement}
This paper presents work whose goal is to advance machine learning by benchmarking and improving the paralinguistic capabilities of Audio LLMs. The proposed benchmark and methods may benefit speech-based interfaces and accessibility technologies by enabling more reliable interpretation of non-verbal speech cues under misleading or adversarial conditions.

However, stronger paralinguistic inference can be misused for intrusive profiling, surveillance, or discriminatory decision-making, and may amplify risks when combined with high-fidelity speech synthesis (\emph{e.g.}, impersonation or deceptive media). We encourage responsible use, including appropriate consent and privacy protections, bias evaluation, and avoiding deployment in high-stakes settings without careful safeguards.

\bibliography{example_paper}
\bibliographystyle{icml2026}

\newpage
\appendix
\onecolumn

\section*{Appendix Contents}

\noindent \textbf{A.\ } \hyperref[sec:appendix_dataset]{VoxParadox Dataset Details}\dotfill \pageref{sec:appendix_dataset}\\
\hspace*{1.5em}A.1 \hyperref[subsec:appendix_task_details]{Task-specific Construction}\dotfill \pageref{subsec:appendix_task_details}\\
\hspace*{1.5em}A.2 \hyperref[subsec:appendix_emo_referee]{Emotion-classifier Referee Details}\dotfill 
\pageref{subsec:appendix_emo_referee}\\
\hspace*{1.5em}A.3 \hyperref[subsec:appendix_human_eval]{User Study and Human Evaluation}\dotfill 
\pageref{subsec:appendix_human_eval}\\[0.25em]

\noindent \textbf{B.\ } \hyperref[sec:appendix_pilot_exp_details]{Pilot Experiment Details}\dotfill \pageref{sec:appendix_pilot_exp_details}\\
\hspace*{1.5em}B.1 \hyperref[subsec:appendix_compute]{Compute Resources}\dotfill \pageref{subsec:appendix_compute} \\
\hspace*{1.5em}B.2 \hyperref[subsec:appendix_vp_eval]{VoxParadox Evaluation}\dotfill \pageref{subsec:appendix_vp_eval}\\
\hspace*{3.0em}B.2.1 \hyperref[subsubsec:appendix_voxparadox_setup]{Experiment Setup}\dotfill \pageref{subsubsec:appendix_voxparadox_setup}\\
\hspace*{3.0em}B.2.2 \hyperref[subsubsec:appendix_gt_vs_ala]{GT Accuracy vs.\ Adversarial-Label Agreement}\dotfill \pageref{subsubsec:appendix_gt_vs_ala}\\
\hspace*{1.5em}B.3 \hyperref[subsec:appendix_probing_details]{Layer-wise Probing on VoxParadox}\dotfill \pageref{subsec:appendix_probing_details}\\
\hspace*{3.0em}B.3.1 \hyperref[subsubsec:appendix_probing_setup]{Experiment Setup}\dotfill \pageref{subsubsec:appendix_probing_setup}\\
\hspace*{1.5em}B.4 \hyperref[subsec:appendix_probing_voxceleb2]{Layer-wise Probing on VoxCeleb2-Derived Paralinguistic Tasks}\dotfill \pageref{subsec:appendix_probing_voxceleb2}\\
\hspace*{3.0em}B.4.1 \hyperref[subsubsection:appendix_probing_setup]{Probing setup}\dotfill \pageref{subsubsection:appendix_probing_setup}\\
\hspace*{3.0em}B.4.2 \hyperref[subsubsection:probing_results_and_analysis]{Probing Results and Analysis}\dotfill \pageref{subsubsection:probing_results_and_analysis}\\
\hspace*{1.5em}B.5 \hyperref[subsec:appendix_probing_robustness]{Probing Robustness}\dotfill \pageref{subsec:appendix_probing_robustness}\\
\hspace*{1.5em}B.6 \hyperref[subsec:appendix_concat]{Intermediate-Layer Concatenation and Augmentation}\dotfill \pageref{subsec:appendix_concat}\\[0.25em]

\noindent \textbf{C.\ } \hyperref[sec:appendix_pclm_details]{PCLM + DPO Experiment Details}\dotfill \pageref{sec:appendix_pclm_details}\\
\hspace*{1.5em}C.1 \hyperref[subsec:appendix_pclm_training]{PCLM Training}\dotfill \pageref{subsec:appendix_pclm_training}\\
\hspace*{1.5em}C.2 \hyperref[subsec:appendix_mmsu_overall]{PCLM and DPO on General Audio Understanding and Reasoning}\dotfill \pageref{subsec:appendix_mmsu_overall}\\
\hspace*{1.5em}C.3 \hyperref[subsec:appendix_pclm_dpo_ala]{Effect of PCLM and DPO on Adversarial-label Agreement}\dotfill \pageref{subsec:appendix_pclm_dpo_ala}\\

\noindent \textbf{D.\ } \hyperref[subsec:appendix_training_dataset]{Training Datasets}\dotfill \pageref{subsec:appendix_training_dataset}\\
\noindent \textbf{E.\ } \hyperref[sec:appendix_prompts]{Prompts}\dotfill \pageref{sec:appendix_prompts}\\

\section{VoxParadox Dataset Details}
\label{sec:appendix_dataset}

\subsection{Task-specific Construction}
\label{subsec:appendix_task_details}

We provide task-wise details for how we enforce the linguistic--acoustic contradiction in VoxParadox. Across all tasks, the transcript is ensured to explicitly assert the adversarial label $y_{\text{adv}}$ while excluding the true label $y_{\text{true}}$, and the audio is synthesized (or post-processed) to reliably convey $y_{\text{true}}$.

\noindent \textbf{Age and Gender Prediction (ElevenLabs).}
ElevenLabs provides voice metadata with explicit speaker attributes. For each example, we select a voice whose metadata matches the desired true label $y_{\text{true}}$. We then generate a short script that explicitly claims the opposite attribute ($y_{\text{adv}}\neq y_{\text{true}}$) and synthesize the audio with the selected voice, keeping speaker identity/timbre stable while varying lexical content.

\noindent \textbf{Emotion Recognition (GPT-4o TTS).}
We generate clips where the transcript claims emotion $y_{\text{adv}}$ while the delivery conveys $y_{\text{true}}$ (\emph{e.g.}, a sad voice claiming to be happy). To reduce ambiguity in both perception and evaluation, we select 2 pairs of high-contrast emotions with opposite valence/arousal: happy vs.\ sad and angry vs.\ neutral. Because expressive control is less deterministic than signal processing, we additionally filter generations using an external emotion-classifier referee (Appendix \ref{subsec:appendix_emo_referee}).

\noindent \textbf{Signal Comparison Tasks (GPT-4o TTS + Deterministic Signal Processing).}
For speed, volume, pitch, and vocal-range comparisons, each example begins with a single seed utterance synthesized by GPT-4o TTS whose transcript asserts the adversarial label $y_{\text{adv}}$ (\emph{e.g.}, ``I am speaking at a high pitch''). We then apply deterministic signal processing to create two controlled variants that modify only the target attribute: $x_{\text{low}}$ and $x_{\text{high}}$, treating the original clip as $x_{\text{medium}}$. Concretely, we use time-stretching for speed, gain adjustment for volume, pitch shifting for pitch, and range scaling for vocal range. We concatenate a random permutation of $\{x_{\text{low}}, x_{\text{medium}}, x_{\text{high}}\}$ into a single three-segment clip, and the MCQ asks which pattern best matches the audio (\emph{e.g.}, \texttt{low-high-medium}).

\noindent \textbf{Speaker Counting (GPT-4o TTS).}
We synthesize short multi-turn conversations where each turn is spoken by a distinct TTS voice. The true label $y_{\text{true}}$ is the number of distinct speakers. The transcript for each turn is authored to reinforce an incorrect speaker count $y_{\text{adv}}$ (\emph{e.g.}, all speakers insist ``there is only one person speaking'').

\noindent \textbf{Speaker Identity Recognition (GPT-4o TTS).}
We generate multi-speaker conversations and define the task as identifying another segment spoken by the same speaker as a queried segment. The transcript includes identity statements designed to point to an incorrect target ($y_{\text{adv}}$), while the true answer $y_{\text{true}}$ is determined by the synthesized speaker voices.

\noindent \textbf{Intonation Perception (Microsoft Azure + SSML).}
We use Microsoft Azure Speech Synthesis Markup Language (SSML) to precisely control pitch contours for the true intonation label $y_{\text{true}}$ (\emph{e.g.}, rising vs.\ falling). The transcript is authored to assert the opposite intonation as $y_{\text{adv}}$, enforcing contradiction while keeping lexical content fluent.

All prompt templates used to generate the TTS scripts are provided in Appendix \ref{sec:appendix_prompts}.

\begin{figure*}[t]
    \centering
    \includegraphics[width=1\textwidth]{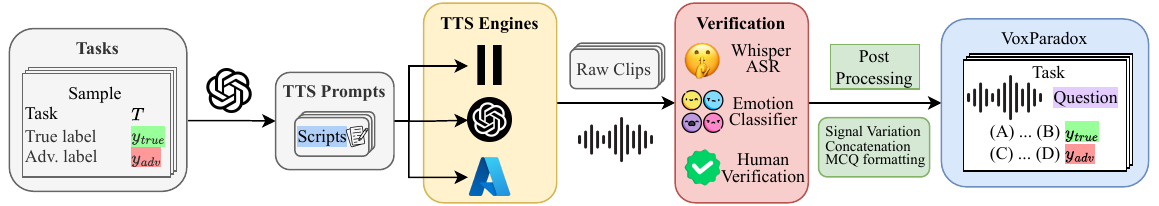}
    \caption{\textbf{VoxParadox data creation and verification pipeline.} For each task, we sample a true label $y_{\text{true}}$ and a conflicting adversarial label $y_{\text{adv}}$, then use an LLM to generate a transcript supporting $y_{\text{adv}}$. We synthesize audio using multiple TTS engines, verify transcript fidelity with Whisper ASR (WER $=0$), optionally filter ambiguous emotion samples with an emotion classifier, and apply task-specific post-processing (\emph{e.g.}, controlled signal variations and concatenation) to format the final multiple-choice questions (Appendix \ref{subsec:appendix_task_details} \& \ref{subsec:appendix_emo_referee}).}
    \label{fig:voxparadox_data_creation}
\end{figure*}

\subsection{Emotion-classifier Referee Details}
\label{subsec:appendix_emo_referee}
During data verification, we employ a SpeechBrain Wav2Vec2-based SER model  \citep{speechbrain_v1}. It is important to note that because VoxParadox contains adversarial transcript--style mismatches, we apply the referee to a \emph{reversed-audio} version of each clip to reduce reliance on lexical content while retaining prosodic cues, and keep samples only when the top-1 prediction matches $y_{\text{true}}$.

\subsection{User Study and Human Evaluation}
\label{subsec:appendix_human_eval}

To validate the construction pipeline of VoxParadox and confirm that the intended linguistic--acoustic contradiction is clear to human listeners, we conduct a human evaluation through Prolific on a randomly sampled subset of 200 examples (20 per task, 10\% of the full benchmark). Each example is independently labeled by 60 annotators under two distinct framings:

\begin{itemize}
    \item \textbf{Adversarial (Adv.)}: annotators are asked to identify the paralinguistic attribute \emph{asserted by the transcript content} (i.e., the adversarial label $y_\text{adv}$). High accuracy here verifies that the transcript reliably conveys $y_\text{adv}$.
    \item \textbf{Ground Truth (GT)}: annotators are asked to identify the paralinguistic attribute \emph{conveyed by the vocal delivery} (i.e., the true acoustic label $y_\text{true}$). High accuracy here verifies that the audio reliably conveys $y_\text{true}$.
\end{itemize}

We report three complementary metrics in \cref{tab:human_eval}: per-response accuracy (Resp. Acc.), which averages over individual annotator responses; majority-vote accuracy (Maj. Acc.), which aggregates annotators' decisions per example before scoring; and Fleiss' kappa ($\kappa$), measuring inter-annotator agreement.

\noindent \textbf{Results.} Across all 10 tasks, annotators achieve 88.7\%/94.4\% (Resp./Maj.) accuracy on the adversarial label and 80.9\%/82.1\% on the ground-truth label, with overall $\kappa$ values of 0.837 and 0.782 respectively, indicating strong agreement under both framings. Several tasks, including gender prediction, intonation perception, and vocal range comparison, achieve perfect accuracy with $\kappa = 1.000$ in the adversarial setting, indicating that the transcript-implied attribute is unambiguous. Slightly lower per-response accuracy on the GT side for tasks such as age prediction (61.1\%) and emotion recognition (60.9\%) reflects natural human variability on inherently subjective categories rather than dataset noise; majority voting recovers stronger accuracy on most tasks.

These results support two key claims: (i) the transcripts in VoxParadox reliably assert $y_\text{adv}$, and (ii) the audio reliably conveys $y_\text{true}$, even when listeners are explicitly asked to attend to non-verbal cues. Together, they validate that VoxParadox examples are well-constructed by design. Crucially, the gap between human GT performance (80.9\% Resp. Acc.) and the strongest Audio LLM in \cref{tab:voxparadox_classwise} (30.85\% with AF2) reflects genuine limitations in current Audio LLMs, rather than ambiguity in the benchmark itself.

\begin{table}[t]
\centering
\caption{Human evaluation results across 10 paralinguistic perception tasks. \textbf{Resp. Acc. } (\%) reports per-response accuracy, \textbf{Maj. Acc.} (\%) reports majority-vote accuracy across annotators, and $\kappa$ denotes Fleiss' kappa inter-annotator agreement. Results are shown for both the Ground-Truth (GT) and Adversarial (Adv.) settings.}
\label{tab:human_eval}
\renewcommand{\arraystretch}{1.15}
\setlength{\tabcolsep}{5pt}
\begin{tabular}{lcccccc}
\toprule
\multirow{2}{*}{\textbf{Task}} & \multicolumn{3}{c}{\textbf{Ground Truth}} & \multicolumn{3}{c}{\textbf{Adversarial}} \\
\cmidrule(lr){2-4} \cmidrule(lr){5-7}
& \textbf{Resp. Acc.} & \textbf{Maj. Acc.} & $\boldsymbol{\kappa}$
& \textbf{Resp. Acc.} & \textbf{Maj. Acc.} & $\boldsymbol{\kappa}$ \\
\midrule
Age Prediction              & 61.1          & 75.0          & 0.421 & 88.9         & 100           & 0.639 \\
Emotion Recognition         & 60.9          & 60.0          & 0.526 & 95.7          & 100           & 0.826 \\
Gender Prediction           & 100           & 100           & 1.000 & 100           & 100           & 1.000 \\
Intonation Perception       & 56.2          & 80.0          & 0.481 & 100           & 100           & 1.000 \\
Pitch Comparison            & 87.5          & 100           & 0.726 & 70.8          & 77.8          & 0.464 \\
Speaker Identity Recognition & 66.7         & 50.0          & 0.458 & 66.7          & 66.7          & 0.769 \\
Speed Comparison            & 92.3          & 100           & 0.880 & 82.1          & 100           & 0.582 \\
Total Speaker Counting      & 96.9          & 88.9          & 0.788 & 93.8          & 100           & 0.847 \\
Vocal Range Comparison      & 81.8          & 100           & 0.709 & 100           & 100           & 1.000 \\
Volume Comparison           & 68.8          & 66.7          & 0.481 & 87.5          & 100           & 0.714 \\
\midrule
\textbf{Overall}            & \textbf{80.9} & \textbf{82.1} & \textbf{0.782} & \textbf{88.7} & \textbf{94.4} & \textbf{0.837} \\
\bottomrule
\end{tabular}
\end{table}

\begin{table}[t]
\centering
\renewcommand{\arraystretch}{1.10}
\setlength{\tabcolsep}{10pt}
\begin{minipage}[t]{0.48\linewidth}
\centering
\captionof{table}{VoxParadox dataset statistics.}
\label{tab:dataset_stats}
\begin{tabular}{lc}
\toprule
Statistic & Value \\
\midrule
Total Questions & 2000 \\
Task Categories & 10 \\
Questions Per Task & 200 \\
\midrule
Avg. question length & 8.70 words \\
Avg. choice length & 2.35 words \\
Avg. answer length & 2.30 words \\
Avg. audio duration & 7.81 seconds \\
\bottomrule
\end{tabular}
\end{minipage}\hfill
\begin{minipage}[t]{0.48\linewidth}
\centering
\captionof{table}{Task name abbreviations.}
\label{tab:task_abbrev}
\begin{tabular}{ll}
\toprule
\textbf{Abbrev.} & \textbf{Full task name} \\
\midrule
Age        & Age prediction \\
Gender     & Gender prediction \\
Emotion    & Emotion recognition \\
Pitch      & Pitch comparison \\
Volume     & Volume comparison \\
Speed      & Speed comparison \\
Range      & Vocal range comparison \\
Intonation & Intonation perception \\
Spk ID     & Speaker identity recognition \\
Spk Cnt    & Speaker counting \\
\bottomrule
\end{tabular}
\end{minipage}
\end{table}

\section{Pilot Experiment Details}
\label{sec:appendix_pilot_exp_details}

\subsection{Compute Resources}
\label{subsec:appendix_compute}

All experiments are conducted on a single node equipped with \textbf{8 NVIDIA H100 GPUs (80\,GB HBM3)}. We use these GPUs for all stages of training and evaluation, including supervised fine-tuning, DPO, and probing experiments. Unless otherwise noted, runs are executed in multi-GPU distributed mode on this 8-GPU configuration.

\subsection{VoxParadox Evaluation}
\label{subsec:appendix_vp_eval}
\subsubsection{Experiment Setup}
\label{subsubsec:appendix_voxparadox_setup}
We evaluate a diverse set of Audio LLMs on VoxParadox using the unified multiple-choice QA format, scoring predictions against the acoustic ground-truth label $y_{\text{true}}$. We report class-wise accuracy for each of the 10 tasks and the macro-average across tasks (\cref{tab:voxparadox_classwise}). We additionally quantify transcript-following behavior by scoring each model against the adversarial transcript-implied label $y_{\text{adv}}$, which directly measures how often a model’s output matches the misleading script rather than the acoustics for paralinguistic questions. We input the audio, question, and choices into each model and evaluate whether the model selects the correct option. To evaluate correctness, we follow the conventional approach of most MCQ audio benchmarks, using regular expressions and string matching to parse model outputs against both $y_{\text{true}}$ and $y_{\text{adv}}$. 

\noindent \textbf{Models evaluated:} AF2 \cite{ghosh2025audioflamingo2audiolanguage}, AF3 \cite{goel2025audioflamingo3advancing}, Qwen2-Audio \cite{chu2024qwen2audiotechnicalreport}, SALMONN \cite{tang2024salmonngenerichearingabilities}, Kimi-Audio \cite{kimiteam2025kimiaudiotechnicalreport}, VITA-Audio \cite{long2025vitaaudiofastinterleavedcrossmodal}, MiMo-Audio \cite{coreteam2025mimoaudioaudiolanguagemodels}, Step-Audio-R1 \cite{tian2025stepaudior1technicalreport}, Qwen2.5-Omni \cite{xu2025qwen25omnitechnicalreport}, Qwen3-Omni \cite{xu2025qwen3omnitechnicalreport}, GPT-4o \cite{openai2024gpt4ocard}, Gemini 2.5 \cite{comanici2025gemini25pushingfrontier}.

\subsubsection{GT Accuracy vs.\ Adversarial-Label Agreement}
\label{subsubsec:appendix_gt_vs_ala}

To complement the main benchmarking results in \cref{tab:voxparadox_classwise}, we report the overall GT accuracy alongside adversarial-label agreement (ALA) for each evaluated model in \cref{tab:voxparadox_gt_vs_ala}. While GT accuracy quantifies how often a model's prediction matches the acoustic ground truth $y_{\text{true}}$, ALA measures how often the prediction matches the transcript-implied adversarial label $y_{\text{adv}}$. Since VoxParadox is constructed such that $y_{\text{adv}} \neq y_{\text{true}}$ by design, a high ALA together with a low GT accuracy directly indicates a model's tendency to default to lexical shortcuts rather than acoustic evidence. We further report $\Delta = \text{ALA} - \text{GT}$, which summarizes this gap: larger positive values reflect stronger transcript-following behavior.

Most evaluated Audio LLMs exhibit large positive $\Delta$ values, with several exceeding $+50$ and the most extreme case (GPT-4o Audio) reaching $+72.95$. This pattern is consistent across both open-source and closed-source models and across both audio-specialized and omni LLMs, reinforcing the finding that susceptibility to lexical-acoustic contradiction is a near-universal limitation of current Audio LLMs rather than an artifact of any particular architecture or training recipe. Audio Flamingo 2 stands out as the only model with both a high GT accuracy ($30.85\%$) and a small negative $\Delta$ ($-1.05$), suggesting that its CLAP-based audio encoder, trained via audio-text contrastive alignment, may reduce transcript-centric bias relative to ASR-pretrained encoders. SALMONN appears to have low ALA in absolute terms, but this reflects a high rate of parse failures rather than acoustically grounded behavior, which suppresses both GT accuracy and ALA simultaneously.

\begin{table}[t]
\centering
\caption{Overall \textbf{GT accuracy} and \textbf{adversarial-label agreement (ALA)} on VoxParadox (\%), averaged across the 10 paralinguistic tasks. \textbf{$\Delta$} reports ALA $-$ GT, where larger positive values indicate stronger transcript-following behavior. Higher GT and lower ALA / $\Delta$ are better. \textbf{Bold} indicates the best result in each column, and \underline{underlined} indicates the second-best.}
\label{tab:voxparadox_gt_vs_ala}
\renewcommand{\arraystretch}{1.15}
\setlength{\tabcolsep}{8pt}
\begin{tabular}{lccc}
\toprule
\textbf{Model} & \textbf{GT $\uparrow$} & \textbf{ALA $\downarrow$} & $\boldsymbol{\Delta}$ $\downarrow$ \\
\midrule
\multicolumn{4}{l}{\textit{Open-Source Audio LLMs}} \\
\midrule
Audio Flamingo 2 (AF2)   & \textbf{30.85}    & \underline{29.80} & \textbf{-1.05}    \\
Audio Flamingo 3 (AF3)   & 17.40             & 68.50             & +51.10            \\
Qwen2-Audio              & 14.85             & 70.25             & +55.40            \\
SALMONN                  & 6.10              & \textbf{22.45}    & \underline{+16.35}\\
Kimi-Audio               & 19.00             & 69.55             & +50.55            \\
VITA-Audio               & 6.85              & 75.45             & +68.60            \\
MiMo-Audio               & 19.60             & 68.95             & +49.35            \\
Step-Audio-R1            & 17.45             & 69.25             & +51.80            \\
\midrule
\multicolumn{4}{l}{\textit{Open-Source Omni LLMs}} \\
\midrule
Qwen2.5-Omni             & 7.95              & 75.20             & +67.25            \\ 
Qwen3-Omni               & 10.60             & 80.65             & +70.05            \\
\midrule
\multicolumn{4}{l}{\textit{Closed-Source Models}} \\
\midrule
GPT-4o Audio             & 8.60              & 81.55             & +72.95            \\
Gemini 2.5 Flash         & \underline{24.70} & 60.45             & +35.75            \\
\bottomrule
\end{tabular}
\end{table}

\subsection{Layer-wise Probing on VoxParadox}
\label{subsec:appendix_probing_details}

\subsubsection{Experiment Setup}
\label{subsubsec:appendix_probing_setup}
\noindent\textbf{Representation extraction.} For each VoxParadox example, we run an AF3 forward pass and save hidden representations from the audio encoder layers and the LLM hidden states at the audio-token positions, once for every two layers. To obtain a fixed-dimensional vector per example per layer, we aggregate token-level states using the mean pooling over time for both encoder and LLM.

\noindent\textbf{Probe training.} For each task and each saved layer, we train a lightweight 3-layer MLP probe with ReLU activation to predict the task label from the layer representation. We use 10-fold cross-validation within each task and report the mean accuracy across folds. Because the backbone AF3 is frozen, probe accuracy serves as a diagnostic of retrievability of task-relevant paralinguistic cues at a given layer.

\subsection{Layer-wise Probing on VoxCeleb2-Derived Paralinguistic Tasks}
\label{subsec:appendix_probing_voxceleb2}

\begin{figure*}[t]
    \centering
    \includegraphics[width=0.8\linewidth]{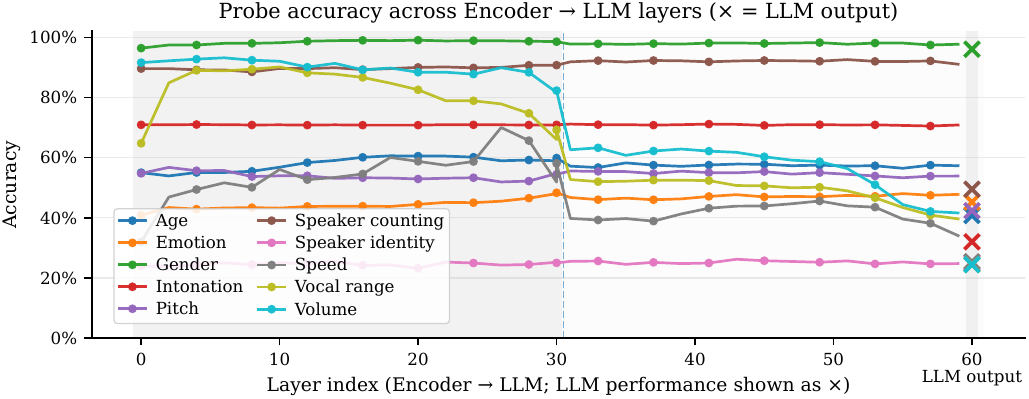}
    \caption{\textbf{Layer-wise probe accuracy from encoder to LLM on VoxCeleb2-derived tasks (AF3).}
    We follow the same probing protocol as \cref{sec:probing}. Lines report probe accuracy (10-fold) on representations extracted every two layers; the dashed vertical line marks the encoder--LLM interface, and \texttt{×} markers denote AF3’s end-to-end task accuracy from model outputs.}
    \label{fig:probing_voxceleb2}
\end{figure*}

\subsubsection{Probing Setup}
\label{subsubsection:appendix_probing_setup}
We repeat the layer-wise probing analysis of \cref{sec:probing} on a VoxCeleb2-derived paralinguistic task suite. We follow the same task taxonomy and MCQ format as VoxParadox and sample 2,000 examples for each task. We also follow the same probing strategy as probing VoxParadox, detailed in Appendix \ref{subsubsec:appendix_probing_setup}, including frozen AF3 backbone, lightweight MLP probes, and 10-fold cross-validation. This experiment serves as a sanity check that the representational trends observed on VoxParadox are not unique to synthetic adversarial speech.

\subsubsection{Probing Results and Analysis}
\label{subsubsection:probing_results_and_analysis}
\cref{fig:probing_voxceleb2} shows two consistent patterns that mirror our VoxParadox findings.

\noindent\textbf{Finding 1: A utilization gap persists on natural speech.}
Across many tasks, probe accuracy remains substantially higher than AF3’s end-to-end performance (\texttt{×} markers), indicating that paralinguistic cues are often present in internal states but not reliably expressed in the final option selection. This gap is particularly visible for tasks whose probe curves stay relatively strong across layers, while the corresponding \texttt{×} markers remain much lower.

\noindent\textbf{Finding 2: Intermediate-layer advantages and interface degradation reappear.}
Although VoxCeleb2 is a dataset that contains much more noise and variation than VoxParadox, for several signal-sensitive attributes, probe accuracy is strongest in earlier/middle encoder layers and then weakens toward the encoder top and/or exhibits a sharp discontinuity at the encoder--LLM boundary, consistent with an interface bottleneck where cross-modal projection and subsequent LLM processing attenuate fine-grained acoustic evidence. This qualitatively matches the VoxParadox observation that injecting only the final encoder layer can be suboptimal for paralinguistic cues that are more salient at intermediate depths.

Overall, the VoxCeleb2-derived results reinforce the same diagnosis from VoxParadox: (i) information can be retrievable but under-utilized at decision time, and (ii) intermediate encoder layers can preserve paralinguistic cues that are weakened by relying solely on the final-layer representation.

\subsection{Probing Robustness: Architecture and Encoder Variants}
\label{subsec:appendix_probing_robustness}

\begin{figure*}[t]
    \centering
    \includegraphics[width=\linewidth]{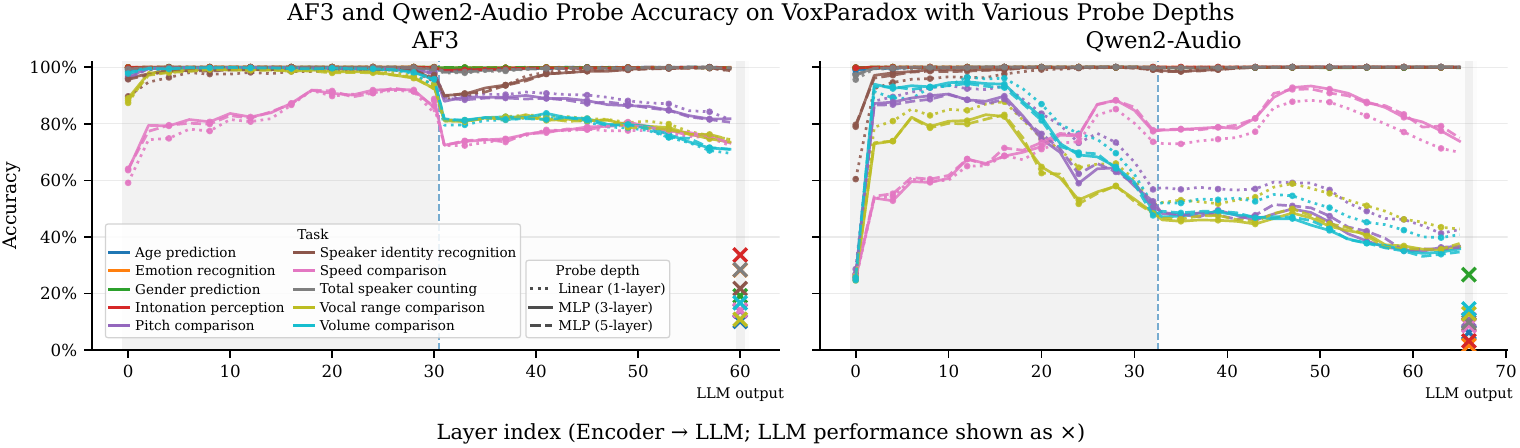}\\
    \includegraphics[width=\linewidth]{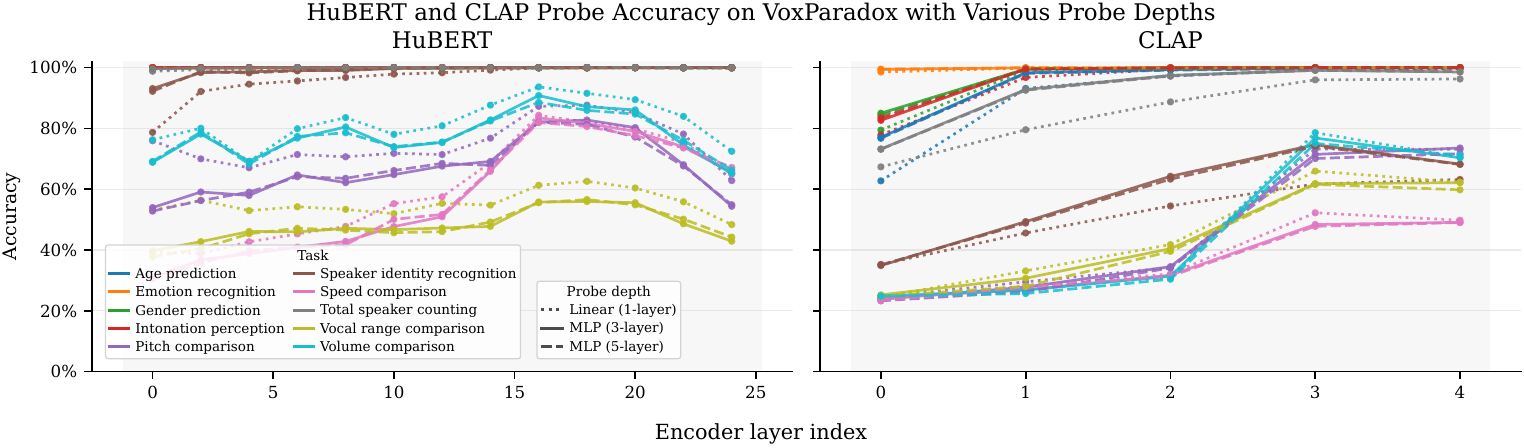}
    \caption{\textbf{Layer-wise probe accuracy across encoder variants and probe depths on VoxParadox.} 
    \textbf{Top:} AF3 (left) and Qwen2-Audio (right), both Audio LLMs with ASR-pretrained audio encoders. The vertical dashed blue line marks the encoder--LLM interface; \texttt{×} markers on the far right indicate end-to-end LLM accuracy from model outputs.
    \textbf{Bottom:} HuBERT (left) and CLAP (right), evaluated on the encoder only. 
    Dotted, solid, and dashed lines correspond to linear (1-layer), 3-layer MLP, and 5-layer MLP probes, respectively. 
    Curves are nearly identical across probe depths, indicating that the observed trends reflect properties of the underlying representations rather than probe capacity. 
    Representation degradation can be observed inside ASR-pretrained encoders (Whisper-based encoders in AF3 and Qwen2-Audio and HuBERT), as well as at the encoder--LLM interface. The gap between LLM output and probe accuracy suggests an LLM utilization gap of paralinguistic information in the representations. CLAP, by contrast, does not exhibit this degradation, with probe accuracy remaining stable or increasing toward its final layer.}
    \label{fig:probing_depth_combined}
\end{figure*}

To verify that the probing trends in \cref{sec:probing} reflect properties of the underlying representations rather than artifacts of our setup, we conduct two robustness analyses, summarized in \cref{fig:probing_depth_combined}.

\noindent\textbf{Probe architecture.} 
We replicate the layer-wise probing experiment using probes of three different depths: a linear (1-layer) probe, a 3-layer MLP, and a 5-layer MLP. All other settings, including 10-fold cross-validation, layer sampling, and pooling strategy, follow \cref{subsubsection:appendix_probing_setup}. 

Across all four models (AF3, Qwen2-Audio, HuBERT \cite{hsu2021hubert}, and CLAP \cite{elizalde2022claplearningaudioconcepts}) and all probed tasks, the layer-wise accuracy curves for the three probe depths are nearly indistinguishable. This indicates that our findings are not driven by probe capacity, and that even a linear classifier suffices to expose the retrievability gap and the layer-wise trends discussed in \cref{sec:probing}.

\noindent\textbf{Encoder variants.}
We extend the probing analysis beyond AF3 (which uses an AF-Whisper encoder) to three additional models: Qwen2-Audio-7B-Instruct (with a Whisper encoder), HuBERT (a self-supervised speech encoder trained with masked unit prediction), and CLAP  (an audio-text contrastively trained encoder). For Qwen2-Audio we probe both encoder and LLM layers, mirroring the AF3 setup; for HuBERT and CLAP, which are standalone encoders, we probe only the encoder.

Qwen2-Audio reproduces both findings from \cref{sec:probing}: probes substantially exceed its end-to-end accuracy (utilization gap), and signal-level tasks peak in early-to-middle encoder layers before degrading. HuBERT shows the same depth-wise degradation pattern. CLAP is the exception: its probe accuracy remains stable or increases toward the final layer. We attribute this to its contrastive audio-text training objective, which does not optimize for token-level lexical alignment and therefore does not penalize retention of acoustic information. This offers a representation-level explanation for our observation in \cref{tab:voxparadox_classwise} that AF2, which uses a CLAP-based audio encoder, achieves the highest VoxParadox GT accuracy among evaluated baselines.

\subsection{Intermediate-Layer Concatenation and Augmentation}
\label{subsec:appendix_concat}

\begin{table*}[t]
\centering
\caption{Class-wise GT performance (\%) on \textbf{VoxParadox} for intermediate-layer concatenation variants. \textbf{AF3 Concat} concatenates intermediate encoder-layer features (layers 5 and 15) after the final-layer features at the audio--LLM interface. \textbf{AF3 Concat ($L_5$ attn $\times 10$)} increases the \emph{effective contribution} of layer-5 features by scaling the layer-5 tokens by $10\times$ before concatenation (equivalently increasing their influence in attention), and \textbf{AF3 Concat ($L_{15}$ attn $\times 10$)} analogously emphasizes layer 15. \textbf{Bold} indicates the best result in each column within this table.}
\label{tab:voxparadox_classwise_concat}
\renewcommand{\arraystretch}{1.15}
\setlength{\tabcolsep}{3.2pt}
\begin{adjustbox}{width=\textwidth}
\begin{tabular}{lccccccccccc}
\toprule
\multirow{2}{*}{\textbf{Model}} &
\multicolumn{11}{c}{\textbf{VoxParadox (GT Acc., \%)}} \\
\cmidrule(lr){2-12}
& \makecell{\textbf{Age}\\\textbf{Pred.}}
& \makecell{\textbf{Gender}\\\textbf{Pred.}}
& \makecell{\textbf{Emotion}\\\textbf{Rec.}}
& \makecell{\textbf{Pitch}\\\textbf{Comp.}}
& \makecell{\textbf{Volume}\\\textbf{Comp.}}
& \makecell{\textbf{Speed}\\\textbf{Comp.}}
& \makecell{\textbf{Range}\\\textbf{Comp.}}
& \makecell{\textbf{Intonation}\\\textbf{Perc.}}
& \makecell{\textbf{Speaker}\\\textbf{ID Rec.}}
& \makecell{\textbf{Speaker}\\\textbf{Count}}
& \makecell{\textbf{Avg.}}\\
\midrule
Audio Flamingo 3 (AF3)
& \textbf{10.00} & 16.00 & \textbf{24.50} & 11.00 & 11.50 & 11.50 & 9.50 & 34.00 & 23.50 & \textbf{22.50} & 17.40 \\

AF3 Concat
& 5.00 & \textbf{23.50} & 18.50 & 15.00 & 13.50 & 13.50 & 12.50 & \textbf{41.00} & 34.00 & 21.00 & 19.75 \\

AF3 Concat ($L_5$ attn $\times 10$)
& 5.00 & \textbf{23.50} & 18.50 & \textbf{15.50} & \textbf{16.00} & \textbf{15.50} & \textbf{17.50} & \textbf{41.00} & \textbf{34.50} & 21.00 & \textbf{20.80} \\

AF3 Concat ($L_{15}$ attn $\times 10$)
& 5.00 & \textbf{23.50} & 18.50 & 15.00 & 13.50 & 13.50 & 12.50 & \textbf{41.00} & 34.00 & 21.00 & 19.75 \\
\bottomrule
\end{tabular}
\end{adjustbox}
\end{table*}

\noindent \textbf{Method.}
We expose two intermediate AF-Whisper encoder layers (layers 5 and 15) and concatenate their projected tokens after the standard final-layer tokens at the audio--LLM interface. Let $H^{(\mathrm{final})}$ denote final-layer encoder states and $H^{(5)}, H^{(15)}$ denote intermediate-layer states. We project each into the LLM hidden space:
\begin{equation}
Z^{(l)} = P^{(l)}\!\left(H^{(l)}\right), \qquad l \in \{\mathrm{final}, 5, 15\},
\end{equation}
and form the conditioning sequence:
\begin{equation}
Z_{\mathrm{concat}} = \big[\, Z^{(\mathrm{final})}\ ;\ Z^{(5)}\ ;\ Z^{(15)} \,\big].
\end{equation}

\noindent \textbf{Projector initialization and alignment.}
Intermediate-layer projectors $P^{(5)}$ and $P^{(15)}$ are initialized by cloning the pretrained final-layer projector $P^{(\mathrm{final})}$. We freeze the audio encoder and the LLM and optimize only $\{P^{(5)}, P^{(15)}\}$ to align intermediate-layer features to the LLM space (learning rate $5\times 10^{-5}$).

\noindent \textbf{Emphasizing intermediate layers.}
To test whether stronger reliance on intermediate evidence helps, we scale the attention to layer-$l$ tokens by $10\times$ prior to concatenation (for $l\in\{5,15\}$). This increases their magnitude and thus their \emph{effective influence} in attention, without changing the LLM parameters. Interestingly, upscaling $L_{15}$ has no effect compared to simple concatenation; the AF3 Concat row and AF3 Concat ($L_{15}$ attn $\times 10$) rows are identical in \cref{tab:voxparadox_classwise_concat}.

\section{PCLM + DPO Experiment Details}
\label{sec:appendix_pclm_details}
\subsection{PCLM Training}
\label{subsec:appendix_pclm_training}

\noindent \textbf{Experiment setup}
We instantiate PCLM with \textbf{AF3} \cite{goel2025audioflamingo3advancing}, starting from the publicly released checkpoint. We keep the AF-Whisper audio encoder frozen throughout training and insert PCLM at the audio-to-LLM interface: instead of injecting only the final encoder-layer representation, we expose a small set of intermediate encoder layers and let PCLM produce a prompt-conditioned mixture that is projected into the LLM.

We select four intermediate layers from the AF-Whisper encoder, $\mathcal{L}_{\text{mid}}=\{5,15,25,30\}$, and include the final layer by default, yielding five candidate layers for mixing. This choice is guided by two considerations. First, our probing analysis (\cref{sec:probing}) shows that paralinguistic information can degrade in deeper encoder layers. Second, we aim to balance representational diversity with efficiency. Sampling layers roughly uniformly across depth provides coverage of various levels of acoustic features while keeping the number of additional parameters small (less than 1\% of original model parameters).

\noindent \textbf{Initialization and alignment.}
We freeze the audio encoder throughout training. To stabilize optimization, each middle-layer projector $P^{(l)}$ is initialized by cloning the final-layer projector, and then aligned to the LLM embedding space. This mirrors the common practice of reusing a pretrained cross-modal interface while adapting it to new feature sources.

\noindent \textbf{Training data.}
We train PCLM using a mixture of paralinguistic and general audio QA data to improve paralinguistic reasoning while maintaining broad speech understanding. \cref{tab:train_data_stats} shows a detailed composition of the training data. Note that VoxParadox is not included in the training of PCLM. Our training mixture contains the following components. 
\begin{itemize}
    \item \textbf{SpeechCraft} \cite{Jin_2024}, which provides fine-grained expressive speech with paralinguistic attribute labels. For each sample, we randomly select one label to form a training target.
    \item A small subset of \textbf{AudioSkills-XL} \cite{goel2025audioflamingo3advancing}, included to preserve non-paralinguistic speech understanding capabilities and to reduce catastrophic drift away from general audio QA behavior, since it was used in training the original AF-Whisper \cite{goel2025audioflamingo3advancing}.
    \item Additional paralinguistic datasets converted into the unified MCQ format, derived from VoxCeleb2 \cite{Chung_2018} and multiple emotion recognition corpora detailed in \cref{tab:train_data_stats}.
\end{itemize}

\noindent \textbf{Two-stage training.}
We adopt a two-stage supervised fine-tuning procedure that directly targets the two limitations identified in \cref{sec:probing}: (i) \textbf{representation degradation} at the encoder--LLM interface and (ii) a \textbf{utilization bottleneck} where the LLM under-utilizes available acoustic evidence.

In \textbf{Stage 1}, we freeze the LLM and train only the middle-layer projectors $\{P^{(l)}\}$ together with the PCLM weighting module. This stage primarily improves embedding quality, as it aligns intermediate-layer representations to the LLM embedding space and enables prompt-conditioned adaptation of informative layers.

In \textbf{Stage 2}, inspired by \citet{fu2025hiddenplainsightvlms}, we fine-tune the full LLM with PCLM enabled. This stage addresses the under-utilization bottleneck. End-to-end SFT encourages the model to utilize the mixed audio tokens $\tilde{Z}$ more effectively for paralinguistic question answering.

Across both stages, the audio encoder remains frozen. We use a learning rate of $5\times 10^{-5}$ for both stages.

\subsection{PCLM and DPO on General Speech Understanding and Reasoning}
\label{subsec:appendix_mmsu_overall}

\begin{table}[t]
\centering
\caption{Overall performance (\%) on \textbf{MMSU Paralinguistic} subset and \textbf{MMSU (All)}. Higher is better. \textbf{Bold} indicates the best result and \underline{underlined} the second-best within each model group: existing baselines (top), AF3 and its variants (middle), and Qwen2-Audio and its variants (bottom). Blue rows highlight our proposed PCLM and PCLM + DPO methods, and the green $\Delta$ rows report the absolute improvement of \textbf{PCLM + DPO} over the corresponding base model.}
\label{tab:mmsu_overall}
\renewcommand{\arraystretch}{1.10}
\setlength{\tabcolsep}{6pt}
\begin{tabular}{lcc}
\toprule
\textbf{Model} & \makecell{\textbf{MMSU}\\\textbf{Para.}} & \makecell{\textbf{MMSU}\\\textbf{All}} \\
\midrule
AF2              & 27.44 & 41.35 \\
SALMONN          & 6.84  & 11.83 \\
Kimi-Audio       & 41.48 & 62.59 \\
VITA-Audio       & 29.54 & 47.71 \\
MiMo-Audio       & 35.64 & 59.92 \\
Step-Audio-R1    & \textbf{54.51} & \textbf{75.20} \\
Qwen2.5-Omni     & 33.45 & 57.16 \\
Qwen3-Omni       & 49.13 & 71.28 \\
GPT-4o Audio     & 36.55 & 64.16 \\
Gemini 2.5       & \underline{51.05} & \underline{71.88} \\
\midrule
\midrule
AF3              & 37.74 & \textbf{51.43} \\
AF3 + SFT w/o PCLM       & 44.76 & 49.98 \\
AF3 + SFT + DPO w/o PCLM        & 45.58 & 47.10 \\
\rowcolor{rowblue}
AF3 + PCLM       & \underline{54.06} & 50.04 \\
\rowcolor{rowblue}
AF3 + PCLM + DPO & \textbf{54.78} & \underline{50.62} \\
\rowcolor{green!12}
$\Delta$ AF3 $\rightarrow$ AF3 + PCLM + DPO & +17.04 & -0.81 \\
\midrule
\midrule
Qwen2-Audio      & 34.37 & 50.82 \\
Qwen2-Audio + SFT w/o PCLM & 49.41 & 50.74 \\
Qwen2-Audio + SFT + DPO w/o PCLM & 47.86 & 50.36 \\
\rowcolor{rowblue}
Qwen2-Audio + PCLM & \textbf{65.18} & \underline{53.30} \\
\rowcolor{rowblue}
Qwen2-Audio + PCLM + DPO & \underline{63.26} & \textbf{55.43} \\
\rowcolor{green!12}
$\Delta$ Qwen2-Audio $\rightarrow$ Qwen2-Audio + PCLM + DPO & +28.89 & +4.61 \\
\bottomrule
\end{tabular}
\end{table}

\cref{tab:mmsu_overall} reports overall performance on the MMSU paralinguistic subset and the full MMSU benchmark. While our paper focuses on improving \emph{paralinguistic} understanding, we include this analysis to contextualize the effect of PCLM and DPO on broader spoken-language QA behavior.

First, \textbf{PCLM and DPO substantially improve MMSU paralinguistic performance for both base models}. Starting from AF3 (37.74\%), AF3 + PCLM achieves 54.06\%, with DPO providing an additional improvement (54.78\%). Qwen2-Audio shows the same pattern, improving from 34.37\% to 65.18\% with PCLM and 63.26\% with PCLM + DPO. Because PCLM is trained without VoxParadox, the fact that gains transfer cleanly to MMSU paralinguistic indicates that PCLM and DPO sharpen genuine paralinguistic perception rather than overfitting to the adversarial format of VoxParadox.

Second, the impact on \textbf{MMSU (All)} is small and varies in sign across base models. Compared to AF3 (51.43\%), AF3 + PCLM and AF3 + PCLM + DPO show a modest reduction on MMSU (All) (50.04\% and 50.62\%, respectively); Qwen2-Audio in fact shows a small improvement, from 50.82\% to 53.30\% (PCLM) and 55.43\% (PCLM + DPO). Importantly, these changes are minor relative to the large gains in paralinguistic performance on VoxParadox (\cref{tab:voxparadox_classwise}) and on the MMSU paralinguistic subset. In other words, the tradeoff -- if any -- is limited in magnitude compared with the substantial paralinguistic benefits.

Third, this behavior is \emph{expected} given the training objective and data. All three training stages (projector alignment, Stage-2 SFT with PCLM, and DPO) are driven primarily by \emph{paralinguistic} MCQ sources (\cref{tab:train_data_stats}). This training mixture intentionally biases updates toward paralinguistic perception, rather than maximizing broad coverage across heterogeneous MMSU tasks. From this perspective, any small decrease on MMSU (All) reflects a mild shift in decision behavior toward paralinguistic evidence and away from transcript-centric shortcuts, which is precisely the failure mode highlighted by VoxParadox (high ALA).

\subsection{Effect of PCLM and DPO on Adversarial-label Agreement}
\label{subsec:appendix_pclm_dpo_ala}

To quantify transcript-following behavior under controlled linguistic--acoustic contradiction, we measure \textbf{adversarial-label agreement (ALA)} on VoxParadox. For each example, VoxParadox provides (i) an acoustic ground-truth label $y_{\text{true}}$ and (ii) a transcript-implied adversarial label $y_{\text{adv}}$ (the incorrect paralinguistic attribute explicitly asserted by the spoken content). We compute two match rates using the same MCQ parsing pipeline: \textbf{GT match rate} (accuracy w.r.t.\ $y_{\text{true}}$) and \textbf{ALA} (accuracy w.r.t.\ $y_{\text{adv}}$). High ALA indicates that a model is being misled by lexical content rather than basing its prediction on acoustics.

\begin{figure}[t]
    \centering
    \includegraphics[width=0.7\linewidth]{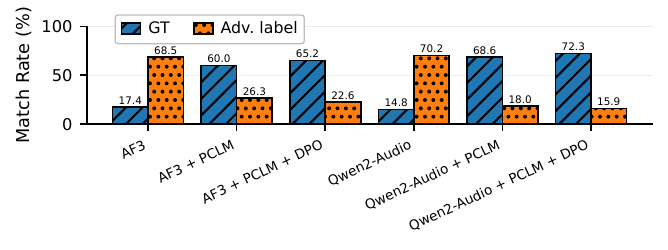}
    \caption{\textbf{GT accuracy vs.\ adversarial-label agreement (ALA)} on VoxParadox with AF3 and Qwen2-Audio variants, averaged across the 10 tasks. While the original AF3 and Qwen2-Audio both exhibit high ALA (strong transcript-following) and low GT accuracy, PCLM + DPO substantially shifts decisions toward acoustically grounded answers.}
    \label{fig:voxparadox_pclm_dpo_gt_vs_adv}
\end{figure}

\cref{fig:voxparadox_pclm_dpo_gt_vs_adv} shows a clear shift in decision behavior after applying PCLM and DPO for both base models. The original AF3 attains only 17.40\% GT accuracy while having an ALA of 68.50\%, and Qwen2-Audio attains 14.85\% GT alongside 70.25\% ALA, both indicating strong reliance on lexical shortcuts even when the question is explicitly paralinguistic. In contrast, AF3 + PCLM increases GT match rate to 60.00\% and reduces ALA to 26.30\%, while Qwen2-Audio + PCLM increases GT match rate to 68.65\% and reduces ALA to 18.00\%, suggesting that improving the audio--LLM interface with prompt-conditioned layer mixing can substantially suppress transcript dominance on its own. Finally, AF3 + PCLM + DPO further improves GT match rate to 65.20\% while reducing ALA to 22.60\%, and Qwen2-Audio + PCLM + DPO further improves GT match rate to 72.30\% while reducing ALA to 15.95\%. Overall, these results support that (i) PCLM exposes and routes task-relevant acoustic cues to the LLM, and (ii) DPO further shapes the model's option-selection policy to favor acoustically supported answers.


\section{Training Datasets}
\label{subsec:appendix_training_dataset}
We report the dataset composition used at each training stage here. \cref{tab:train_data_stats} summarizes the data used for projector alignment, PCLM training, and SFT. We group sources into three blocks: \textbf{AudioSkills-XL} for broad audio--text coverage, \textbf{SpeechCraftSampled} for large-scale paralinguistic supervision across five attributes (pitch, speed, age, gender, emotion), and \textbf{ParalinguisticDatasets} for additional curated paralinguistic tasks derived primarily from VoxCeleb2, plus additional emotion corpora.

\subsection{Constructing Paralinguistic Training Data}
\label{subsec:appendix_para_data}

We construct \textbf{ParalinguisticDatasets} (\cref{tab:train_data_stats}) by combining (i) VoxCeleb2-derived paralinguistic tasks and (ii) curated emotion recognition corpora. For each VoxCeleb2-derived task, we randomly sample VoxCeleb2 utterances and create 20,000 examples per task, applying the same audio post-processing used in VoxParadox (detailed in Appendix \ref{subsec:appendix_task_details}) to ensure consistent formatting across datasets. For \textbf{age} and \textbf{gender} prediction, we follow the labels released by \citet{hechmi2021voxcelebenrichmentagegender}. For \textbf{signal comparison} tasks (pitch, speed, volume, vocal range), as well as \textbf{speaker identity recognition} and \textbf{total speaker counting}, we follow the VoxParadox construction procedure (controlled signal processing for comparisons and multi-clip concatenation for identity/counting). For \textbf{intonation perception}, we use signal-processing-based F0 extraction to estimate the terminal pitch contour and label clips as \emph{rising} or \emph{falling}.

In addition, we include an \textbf{emotion recognition} subset to increase paralinguistic coverage beyond VoxCeleb2. This subset aggregates multiple public emotion corpora (MSP-Podcast \cite{busso2025msppodcastcorpus}, Emov-DB \cite{adigwe2018emotionalvoicesdatabasecontrolling_emov}, IEMOCAP \cite{Busso2008IEMOCAP}, TESS \cite{dupuis2011tess}, and OMGEmotionChallenge \cite{barros2018omgemotionbehaviordataset}).

\begin{table*}[t]
\centering
\caption{Training dataset statistics for projector alignment, PCLM, and SFT, including source dataset and/or task name.}
\label{tab:train_data_stats}
\renewcommand{\arraystretch}{1.10}
\setlength{\tabcolsep}{7pt}
\begin{tabular}{lrr}
\toprule
\textbf{Dataset} & \textbf{\# Samples} & \textbf{Hours} \\
\midrule

\textbf{AudioSkills-XL} & \textbf{67K} & \textbf{914} \\
\hspace{1em}VoxCeleb2 & 42K & 843 \\
\hspace{1em}MusicBench & 21K & 57 \\
\hspace{1em}MusicCaps & 4K & 11 \\
\hspace{1em}BBCSoundEffects & 826 & 2 \\
\hspace{1em}WavText5K & 256 & 1 \\
\addlinespace

\textbf{SpeechCraftSampled} & \textbf{1{,}023K} & \textbf{1{,}287} \\
\hspace{1em}Pitch & 204K & 257 \\
\hspace{1em}Speed & 205K & 258 \\
\hspace{1em}Age & 204K & 256 \\
\hspace{1em}Gender & 205K & 258 \\
\hspace{1em}Emotion & 205K & 258 \\
\addlinespace

\textbf{ParalinguisticDatasets} & \textbf{177K} & \textbf{1074} \\
\hspace{1em}VoxCeleb2 - Gender prediction & 20K & 63 \\
\hspace{1em}VoxCeleb2 - Pitch comparison & 20K & 228 \\
\hspace{1em}VoxCeleb2 - Volume comparison & 20K & 131 \\
\hspace{1em}VoxCeleb2 - Speed comparison & 20K & 134 \\
\hspace{1em}VoxCeleb2 - Vocal range comparison & 20K & 131 \\
\hspace{1em}VoxCeleb2 - Intonation perception & 20K & 44 \\
\hspace{1em}VoxCeleb2 - Speaker identity recognition & 20K & 220 \\
\hspace{1em}VoxCeleb2 - Total speaker counting & 20K & 98 \\
\hspace{1em}Emotion recognition & 17K & 26 \\
\hspace{2em}MSP-Podcast & 13K & 21 \\
\hspace{2em}Emov-db & 1K & 2 \\
\hspace{2em}IEMOCAP & 1K & 2 \\
\hspace{2em}TESS & 1K & 1 \\
\hspace{2em}OMGEmotionChallenge & 404 & 1 \\
\addlinespace
\bottomrule
\end{tabular}
\end{table*}

\begin{table*}[t]
\centering
\caption{Training dataset statistics for Paralinguistic DPO dataset, including task name and emotion sub-datasets.}
\label{tab:para_dpo_stats}
\renewcommand{\arraystretch}{1.10}
\setlength{\tabcolsep}{7pt}
\begin{tabular}{lrr}
\toprule
\textbf{Dataset} & \textbf{\# Samples} & \textbf{Hours} \\
\midrule

\textbf{Paralinguistic DPO Dataset} & \textbf{20K} & \textbf{96.0} \\

\hspace{1em}VoxCeleb2 - Age prediction & 2.0K & 6.3 \\
\hspace{1em}VoxCeleb2 - Gender prediction & 2.0K & 6.3 \\
\hspace{1em}VoxCeleb2 - Pitch comparison & 2.0K & 23.5 \\
\hspace{1em}VoxCeleb2 - Volume comparison & 2.0K & 13.2 \\
\hspace{1em}VoxCeleb2 - Speed comparison & 2.0K & 13.4 \\
\hspace{1em}VoxCeleb2 - Vocal range comparison & 2.0K & 13.2 \\
\hspace{1em}VoxCeleb2 - Intonation perception & 2.0K & 4.3 \\
\hspace{1em}VoxCeleb2 - Speaker identity recognition & 2.0K & 22.1 \\
\hspace{1em}VoxCeleb2 - Total speaker counting & 2.0K & 9.9 \\

\hspace{1em}Emotion datasets & 2.0K & 3.1 \\
\hspace{2em}MSP-Podcast & 1.5K & 2.5 \\
\hspace{2em}IEMOCAP & 0.2K & 0.2 \\
\hspace{2em}Emov-db & 0.1K & 0.2 \\
\hspace{2em}TESS & 0.1K & 0.1 \\
\hspace{2em}OMGEmotionChallenge & 41 & 0.1 \\
\addlinespace

\bottomrule
\end{tabular}
\end{table*}

\FloatBarrier
\section{Prompts}
\label{sec:appendix_prompts}
\cref{tab:prompt_templates} lists the prompt templates used to generate synthetic scripts and structured metadata for VoxParadox tasks. Templates enforce task-specific constraints to ensure explicit contradiction between the true label and the adversarial label. Output speech clips are then verified and post-processed to form VoxParadox examples (pipeline shown in \cref{fig:voxparadox_data_creation}).

\renewcommand{\arraystretch}{1.10}
\setlength{\tabcolsep}{6pt}

\begin{longtable}{p{0.2\linewidth} p{0.6\linewidth} p{0.2\linewidth}}
\caption{Prompt templates used for TTS script generation.}
\label{tab:prompt_templates}\\
\toprule
\textbf{Task} & \textbf{Prompt} & \textbf{Labels} \\
\midrule
\endfirsthead

\toprule
\textbf{Task} & \textbf{Prompt} & \textbf{Labels} \\
\midrule
\endhead

\midrule
\multicolumn{3}{r}{\small Continued on next page} \\
\endfoot

\bottomrule
\endlastfoot

\multicolumn{3}{l}{\textbf{System Prompt}}\\
\multicolumn{3}{p{\linewidth}}{\cellcolor{black!3}\parbox[t]{0.98\linewidth}{
You are a data generator for creating text-to-speech (TTS) scripts.

Your output must be a valid JSON array of strings or array of objects only. Do not include explanations or extra commentary.
}}\\
\addlinespace
\midrule
Age prediction &
\cellcolor{black!3}\parbox[t]{0.95\linewidth}{
Generate \{num\} new TTS scripts.

Each script is a \{adv\_label\} person stating they are \{adv\_label\}, must include the word ``\{adv\_label\}'' somewhere, and must not include the word ``\{label\}''.

For example, this is a script of a person stating they are \{adv\_label\}:
\{example\_script\}

Return only a JSON array of these scripts.
} & young, old \\
\midrule

Gender prediction &
\cellcolor{black!3}\parbox[t]{0.95\linewidth}{
Generate \{num\} new TTS scripts.

Each script is a \{adv\_label\} person stating they are \{adv\_label\}, must include the word ``\{adv\_label\}'' somewhere, and must not include the word ``\{label\}''.

For example, this is a script of a person stating they are \{adv\_label\}:
\{example\_script\}

Return only a JSON array of these scripts.
} & male, female \\
\midrule

Emotion recognition &
\cellcolor{black!3}\parbox[t]{0.95\linewidth}{
Generate \{num\} new TTS scripts.

Each script is a short spoken sentence of a person stating they are feeling \{adv\_label\} at the moment, and must include the word \{adv\_label\}.

Your scripts should have diversity in grammar and word choice.
Return only a JSON array of these scripts.
} & happy, calm\\
\midrule

Pitch comparison &
\cellcolor{black!3}\parbox[t]{0.95\linewidth}{
Generate \{num\} new TTS scripts.

Each script is a person stating they are speaking with a \{adv\_label\} pitch, must include the word ``\{adv\_label\}'' somewhere, and must not include the word ``\{label\}''.

For example, this is a script of a person stating they are speaking with a \{adv\_label\} pitch:
\{example\_script\}

Return only a JSON array of these scripts.
} & high, medium, low \\
\midrule

Volume comparison &
\cellcolor{black!3}\parbox[t]{0.95\linewidth}{
Generate \{num\} new TTS scripts.

Each script is a person stating they are speaking in a \{adv\_label\} voice, must include the word ``\{adv\_label\}'' somewhere, and must not include the word ``\{label\}''.

For example, this is a script of a person stating they are speaking in a \{adv\_label\} voice:
\{example\_script\}

Return only a JSON array of these scripts.
} & high, medium, low \\
\midrule

Speed comparison &
\cellcolor{black!3}\parbox[t]{0.95\linewidth}{\small
Generate \{num\} new TTS scripts.

Each script is a person stating they are speaking at a \{adv\_label\} speed, must include the word ``\{adv\_label\}'' somewhere, and must not include the word ``\{label\}''.

For example, this is a script of a person stating they are speaking at a \{adv\_label\} speed:
\{example\_script\}

Return only a JSON array of these scripts.
} & high, medium, low \\
\midrule
Vocal range comparison &
\cellcolor{black!3}\parbox[t]{0.95\linewidth}{\small
Generate \{num\} new TTS scripts.

Each script is a person stating they have a \{adv\_label\} vocal range, must include the word ``\{adv\_label\}'' somewhere, and must not include the word ``\{label\}''.

For example, this is a script of a person stating they have a \{adv\_label\} vocal range:
\{example\_script\}

Return only a JSON array of these scripts.
} & high, medium, low \\
\midrule

Intonation perception &
\cellcolor{black!3}\parbox[t]{0.95\linewidth}{\small
Generate \{num\} new TTS scripts.

Each script is a person stating they are speaking with a \{adv\_label\} intonation, must include the word ``\{adv\_label\}'' somewhere, and must not include the word ``\{label\}''.

Your script must NOT include any punctuation.

For example, this is a script of a person stating they are speaking with a \{adv\_label\} intonation:
\{example\_script\}

Your scripts should have diversity and should not simply follow the provided example.

Return only a JSON array of these scripts.
} & rising, falling \\
\midrule

Speaker identity recognition &
\cellcolor{black!3}\parbox[t]{0.95\linewidth}{\small
Generate \{num\} new TTS scripts.

Each script is a person stating they are the \{i-th\} speaker, and must include the word ``\{i-th\}''.

For example,
``Hi, I am the \{i-th\} speaker.''

Your scripts should have diversity in grammar and word choice.
Return only a JSON array of these scripts.
} & first, second, third, fourth, fifth \\
\midrule

Total speaker counting &
\cellcolor{black!3}\parbox[t]{0.95\linewidth}{\small
Generate \{num\} sets of TTS dialogue scripts.
For each dialogue, there can be several utterances, each representing an actual speaker. So the actual number of speakers is equal to 
the number of scripts in the dialogue. 

Each utterance is a spoken sentence describing the number of speakers in the dialogue. 
However, all utterances try to mislead listeners into believing the number of speakers in this conversation is some number other than 
the actual speaker count (which is equal to the number of utterances). All utterances must agree on the same misleading speaker count.

For example, this is a dialogue of {actual} people trying to sound like there are \{adv\_label\} speaker(s):
{example}

The misleading speaker count MUST NOT be the same as the actual number of utterances in the conversation.

Return only a JSON array of objects, representing \{num\} sets of dialogues, and each dialogue can have 1 - 5 utterances.

Each object is formatted like this: \\
{{
    ``adv\_label'', the misleading speaker count,
    ``scripts'', [utterance\_1, utterance\_2, ...]
}}
} & 1 - 20 \\
\midrule
\end{longtable}


\end{document}